\documentclass[notitlepage,prd,onecolumn,showpacs,preprintnumbers,nofootinbib,superscriptaddress,floatfix,tightenlines]{revtex4-2}

\usepackage{amsmath,graphicx,float,upgreek,mciteplus,titlesec,appendix}
\renewcommand{\thesection}{\Roman{section}} 
\usepackage{mathtools}
\usepackage{scalerel} 

\titleformat{\section}
[hang]
{\normalsize\bfseries}
{\thesection.}
{1em}
{\MakeUppercase}
\titlespacing{\section}
{10pt}
{15pt}
{8pt}

\titleformat{\subsection}
[hang]
{\bfseries\normalsize\centering}
{\thesubsection.}
{1em}
{}
\titlespacing{\subsection}
{0pt}
{10pt}
{5pt}

\def\eqref#1{{Eq.\!~(\ref{#1})}}
\def\figref#1{{Fig.\!~\ref{#1}}}
\def\secref#1{{Sec.\!~\ref{#1}}}

\usepackage{subfigure}

\usepackage{xcolor}
\definecolor{lcolor}{rgb}{0.5,0,0}
\definecolor{citcolor}{rgb}{0,0.3,0.0}
\usepackage[breaklinks,colorlinks,urlcolor=blue,citecolor=citcolor,linkcolor=lcolor]{hyperref}

\newcommand{\xt}{\mathbf{x}}

\newcommand{\yt}{\mathbf{y}}

\newcommand{\zt}{\mathbf{z}}

\newcommand{\bt}{\mathbf{b}}
\newcommand{\rt}{\mathbf{r}}
\newcommand{\kt}{\mathbf{k}}

\newcommand{\at}{\mathbf{a}}

\newcommand{\Bt}{\mathbf{B}}
\newcommand{\vt}{\mathbf{v}}
\newcommand{\pt}{\mathbf{p}}
\newcommand{\wt}{\mathbf{w}}

\newcommand{\ptbar}{\bar{\mathbf{p}}}
\newcommand{\ktbar}{\bar{\mathbf{k}}}

\newcommand{\sthat}{\hat{\mathbf{s}}}
\newcommand{\tthat}{\hat{\mathbf{t}}}

\newcommand{\ut}{\mathbf{u}}
\newcommand{\utbar}{\bar{\mathbf{u}}}
\newcommand{\st}{\mathbf{s}}
\newcommand{\stbar}{\bar{\mathbf{s}}}
\newcommand{\ttt}{\mathbf{t}}
\newcommand{\ttbar}{\bar{\mathbf{t}}}
\newcommand{\vtbar}{\bar{\mathbf{v}}}
\newcommand{\ztbar}{\bar{\mathbf{z}}}
\newcommand{\wtbar}{\bar{\mathbf{w}}}

\begin{document}

\title{Evolution of eccentricities induced by geometrical and quantum fluctuations in proton-nucleus collisions}

\author{S.~Demirci} 
\affiliation{Department of Physics, P.O.~Box 35, 40014 University of Jyv\"{a}skyl\"{a}, Finland}
\affiliation{Helsinki Institute of Physics, P.O.~Box 64, 00014 University of Helsinki, Finland}

\author{P.\ Guerrero-Rodríguez}
\affiliation{Laboratório de Instrumentação e Física Experimental de Partículas (LIP), Av.\ Professor Gama Pinto 2, 1649-003 Lisboa, Portugal}

\begin{abstract}
We compute the azimuthal eccentricities arising from initial stage fluctuations in high energy proton-nucleus collisions at proper times $\tau \geq 0^+$. We consider two sources of fluctuations, namely the geometrical structure of the proton and the fluctuation of color fields carried by both proton and nucleus. Describing these effects with Gaussian models allows us to analytically calculate the one- and two-point correlators of energy density, from which the eccentricities are obtained. We compute the proper time evolution of these quantities by approximating the Glasma dynamics in terms of linearized Yang-Mills equations, which we solve by assuming free field propagation and adopting the dilute-dense limit.

\end{abstract}
\maketitle

\section{Introduction}

Heavy ion collisions (HICs) allow experimental access to the high energy and density limit of quantum chromodynamics (QCD). Under such extreme conditions, it is believed that a strongly interacting state made up of deconfined quarks and gluons emerges: the quark gluon plasma (QGP). This novel phase of matter is the main object of study of the heavy ion programs at the Relativistic Heavy Ion Collider (RHIC) and the Large Hadron Collider (LHC). Throughout their history, both experiments have provided (and continue to provide) precise measurements hinting at the formation of the QGP and revealing its physical properties (see e.g.\ \cite{Pasechnik:2016wkt} for a review). One of the most notorious examples is given by the measurement of harmonic flow coefficients $\nu_n$, which characterize collective azimuthal correlations of the outgoing particles after a collision event. This observation finds a natural explanation by assuming that the QGP is a low-viscosity fluid, whose expansion (described by quasi-ideal relativistic hydrodynamics) effectively maps the spatial deformation of the initial state of the collision into a momentum anisotropy in the final state.

The fluid paradigm has been highly successful in the analysis of azimuthal correlations in HICs. However, similar correlations were observed also in collisions of smaller-sized and shorter-lived systems, i.e.\ proton-proton (pp) and proton-nucleus (pA) collisions \cite{Schenke:2021mxx,PHENIX:2018lia,ATLAS:2013jmi,CMS:2015yux,CMS:2016fnw,PHENIX:2013ktj}. As such systems were in principle not expected to exhibit collective effects, these measurements opened up an intense discussion on whether the observed flow originates from momentum correlations intrinsic to the colliding objects or from the final state response to the initial geometry of the system. Regarding the contribution that arises from initial geometry, multiple studies \cite{Bzdak:2013zma,Albacete:2016pmp,Albacete:2016gxu,Weller:2017tsr,Mantysaari:2017cni,Moreland:2018jos,Mantysaari:2020axf} suggest that the event-by-event fluctuation of subnucleonic degrees of freedom plays an essential role in the emergence of correlations in small systems. The description of such fluctuations relies heavily on phenomenological models, which include e.g.\ randomly changing partonic structures and/or quantum fluctuations of color charge densities. These are often considered as accessory contributions to the random fluctuation of nucleonic positions, typically modeled in the Glauber picture \cite{Miller:2007ri}. The chosen prescription can be used to characterize the initial shape of the system, either numerically or analytically through the calculation of eccentricities $\varepsilon_n$. For example, in Ref.\ \cite{Demirci:2021kya} a (mostly) analytical hot spot model was applied to describe the initial deformation of pA collision events. Models such as this are typically used to generate initial conditions for the ensuing expansion of the QGP. However, a complete description also requires addressing the evolution of the early stage before reaching the hydrodynamical regime. During this short intermediate phase the system undergoes multiple interactions that lead it towards thermalization, in a process that may as well have a substantial effect over its eccentricity. In the present work we explore this possibility in high energy pA events by combining the initial stage picture presented in Ref.\ \cite{Demirci:2021kya} with a description of the subsequent evolution in terms of free fields, previously applied in Refs.\ \cite{Lappi:2017skr,Guerrero-Rodriguez:2021ask}.

We employ the color glass condensate (CGC) formalism (see e.g.\ \cite{Gelis:2010nm} for a review) as a natural way to describe the partonic structures of proton and nucleus in the high energy limit. In this picture, the density of low-$x$ partons (assumed to be predominantly gluons) is taken to be so large that one can describe them as classical color fields. These fields are generated by the large-$x$ partons, which are represented as an ensemble of static color charges. The local densities of these charges fluctuate according to Gaussian models in both proton and nucleus, as described by the McLerran-Venugopalan (MV) model \cite{McLerran:1993ni,McLerran:1993ka,McLerran:1994vd}. However, their respective spatial configurations are modeled very differently. In the case of the proton, we assume that the large-$x$ partons are concentrated within $N_q$ Gaussianly distributed hot spots of size $r$, with their locations fluctuating on an event-by-event basis within the proton radius $R$ \cite{Demirci:2021kya}. If one assumed the origin of these hot spots to be the valence quarks of the proton, the natural number of them would be $N_q\!=\!3$. In this work, however, we will treat $N_q$ as a free parameter up until the numerical analysis. In any case, the proton is assumed to be dilute on average (although sufficiently dense within each hot spot). This will allow us to approximate operators by the leading term of their expansion in orders of the proton color charge density (weak field approximation). On the other hand, the nucleus is taken to be homogeneous on average, with its color charge density fluctuations included to all orders in the corresponding fields. Also, in order to obtain analytically tractable results, for the nucleus we apply the Golec-Biernat Wusthof (GBW) parametrization \cite{Golec-Biernat:1998zce} of the dipole amplitude, whereas in the case of the proton we use the result derived from the MV model.

We note that many hot spot models with similar parameters have been used previously in the literature. Many of them apply a fully numerical approach, which allows for a more complete treatment of the problem at hand. For example, the IP-Glasma model \cite{Schenke:2012wb,Schenke:2012hg,Mantysaari:2017cni,Mantysaari:2022ypp} uses hot spots to describe various observables. The parameters included in our model and in the IP-Glasma model are mostly the same. However, our model puts some simplifications in place, like the fact that it is formulated in the dilute-dense limit of a pA collision. The nucleus we consider is infinitely large and, on average, homogenous; and does not have nucleon or hot spot fluctuations. Also, our model does not have saturation scale $(Q_s)$ fluctuations as IP-Glasma and the model in Ref.\ \cite{Kumar:2022aly} do. Implementing this in our framework would require us to let the saturation scale of each hot spot fluctuate independently. There are other physics ingredients used in hot spot models that are not included in the model we use, such as the repulsive force between hot spots \cite{Albacete:2016pmp,Mantysaari:2022ffw} and hot spots within hot spots \cite{Kumar:2021zbn}. Another possible extension to our model (and a way of making it more physically accurate) could be considering a fluctuating number of hot spots. This feature would account for large-$x$ partons splitting into more large-$x$ partons, thus generating new hot spots on an event-by-event basis. In this scenario, the computation of physical observables would require performing an additional average over $N_q$. However, for the present work we chose to keep our model as simple and analytically tractable as possible.

In the CGC picture, the evolution of the early pre-thermalization phase (known as Glasma) is described by the classical Yang-Mills (CYM) equations. Unfortunately, these have no general analytical solution. Numerical approaches like the one implemented in the IP-Glasma model rely on lattice calculations initialized at an infinitesimal positive proper time $\tau\!=\!0^+$. On this specific space-time surface it is possible to find an analytical solution to the Yang-Mills equations, which serves as boundary condition to the ensuing evolution. Several methods have been proposed to approach this evolution analytically, such as performing an expansion of the Glasma fields in powers of $\tau$ \cite{Fries:2006pv,Fujii:2008km,Chen:2015wia,Carrington:2020ssh}. However, the resulting series is found to be dominated by UV divergences associated to the spatial derivative, in such a way that each order is more divergent than the previous one. In Ref.\ \cite{Fujii:2008km} it was proposed that these divergences can be effectively resummed into Bessel functions by keeping only the most singular terms of the equations of motion (i.e.\ the higher order derivatives). This turns out to be equivalent to simply linearizing the Yang-Mills equations, an approach later applied in Refs.\ \cite{Lappi:2017skr,Guerrero-Rodriguez:2021ask} to pA and AA collisions, respectively. As will be detailed in \secref{th_f}, in this work we use a similar strategy as \cite{Lappi:2017skr}, where the linearized Yang-Mills equations are solved in the weak field approximation and assuming free field evolution.

We use this framework to compute the energy density one- and two-point functions in a pA collision in the dilute-dense limit at proper times $\tau \geq 0^+$. To do this, we separate the two-point function into four separate terms. These terms contain, respectively: the averages of the fluctuations, the fluctuations on the proton side, the fluctuations on the nucleus side, and the fluctuations from both sides. In a previous calculation performed for $\tau\!=\!0^+$ \cite{Demirci:2021kya}, the combined fluctuation part was argued to be small in comparison to the other contributions and was thus neglected. We assume this is also the case for $\tau\!>\! 0$. We use the remaining correlators to compute the eccentricities at nonzero values of $\tau$. In \cite{Demirci:2021kya} it was shown that the nucleus-side fluctuations give a negligible contribution to the eccentricities in the case of $N_q\!=\!3$ at $\tau\!=\!0^+$ and on these grounds we neglect it as well. Finally, the non-negligible contributions to the energy density two-point function are the average part and the proton fluctuation part. Out of these two only the proton fluctuation part is a source of eccentricity, as the average part of the energy density two-point function is radially symmetric.

The paper is structured as follows. In \secref{th_f} we briefly introduce the models and approximations that we employ in the description of initial stage fluctuations and their $\tau$-evolution. This framework is then applied in \secref{tau_ev} to the analytical calculation of the $\tau$-dependent one- and two-point correlators of the Glasma energy density. In \secref{ecc} these results are used in the computation of eccentricities at some proper time values of $\tau \geq 0^+$. Then, in \secref{fin}, we present our conclusions and discuss potential extensions of this work.

\section{Theoretical framework}\label{th_f}

The characterization of the medium generated in high energy collisions requires us to describe the partonic content of the colliding objects prior to the interaction. Obtaining such a description from first principles has proven elusive so far, because it demands knowledge of physical phenomena at non-perturbative scales. As QCD is still severely limited in this regime, the standard practice is to adopt approximations and models tailored to describe the hadronic structure in specific kinematical limits. In the present work we set out to describe the geometry of the early stage of high energy pA collisions, and thus we will consider the dilute-dense regime. Such an asymmetric situation requires us to model probe and target in different ways, which are briefly discussed in the subsections below. We also devote the last part of this section to introduce the approximations adopted in the description of the subsequent medium evolution.

\subsection{Nucleus: MV model}

Within the CGC framework the nucleus is modeled as a cloud of small-$x$ gluons emitted by a collection of large-$x$ partons. At high energies, the large densities that characterize the small-$x$ degrees of freedom make it appropriate to describe them as classical fields. The dynamics of said fields are encoded in the CYM equations with an external source, which represents the large-$x$ degrees of freedom:
\begin{align}
[D_{\mu},F^{\mu\nu}]=J^{\nu}=\delta^{+\nu}\rho_A^a(\xt,x^-)t^a\approx \delta^{+\nu}\rho_A^a(\xt)\delta(x^-)t^a.\label{ym0}
\end{align}
Here $\rho_A$ is the color charge density of the nucleus and $t^a$ is the $SU(N_c)$ group generator in the fundamental or adjoint representation (for fermionic or gluonic sources, respectively).
The ansatz proposed for the color current $J^{\nu}$ reflects the fact that, due to Lorentz dilation, the large-$x$ degrees of freedom appear to be static with respect to the small-$x$ gluons they radiate. Thus, for a nucleus moving close to the speed of light in the positive $x^3$ direction the color charge density $\rho_A$ is taken as independent of the light-cone time $x^+$. Furthermore, the Lorentz contraction experienced by the ultrarelativistic nucleus motivates the assumption that $\rho_A$ is very close to a delta function in the longitudinal direction $x^-$. We also assume that the large-$x$ partons do not experience any recoil (eikonal approximation), which is reflected in the current having only a $+$ component.

In order to account for the quantum fluctuations of the nuclear wave function, $\rho_A$ is treated as a random variable that changes on an event-by-event basis. These fluctuations are assumed to obey Gaussian statistics, which is the defining feature of the MV model\footnote{This ansatz was devised for large nuclei. In such systems, the color charge distribution is the result of the superposition of a large number of nucleons assumed to be uncorrelated. However, in practice it is often applied to protons at sufficiently high energies.}.
This is also what makes it an analytically tractable model, allowing us to explicitly compute observables as functional averages over the color charge distributions:
\begin{align}
\langle{\cal O}(\rho_A)\rangle=\int [d\rho_A]\exp{\left[-\int d^2\xt\,\frac{\rho^a_A(\xt)\rho^a_A(\xt)}{2\mu^2}\right]}{\cal O}(\rho_A).
\end{align}
The variance of the Gaussian weight, $\mu^2$, is a parameter proportional to the saturation scale of the nucleus, $Q_s^2\!\sim\!g^2\mu^2$. All correlators of color charges are expressed in terms of the following two-point function:
\begin{equation}
\left\langle \rho_A^a(\xt) \rho_A^b(\yt)  \right\rangle = \mu^2\delta^{(2)}(\xt-\yt)\delta^{ab},
\end{equation}
which is the basic building block in this framework. For the purpose of this paper, however, it is more convenient to work directly with the gauge fields generated by these color charges, which we call $\alpha$. By solving \eqref{ym0} in the light cone gauge, said fields are obtained as gauge rotations of the vacuum (i.e.\ pure gauge fields) confined to the plane transverse to the motion of the nucleus \cite{McLerran:1993ka,McLerran:1994vd}:
\begin{align}
\alpha^{i}(\xt)=-\frac{1}{ig}U(\xt)\partial^{i}U^{\dagger}(\xt).\label{gfield0}
\end{align}
Here $U$ is the Wilson line, an SU($N_c$) element which encodes the interaction between an external probe with the gluon field in the eikonal approximation:
\begin{align}
U(\xt)={\cal P}^-\exp\left[ -ig\int^{\infty}_{-\infty}dz^-\int d^2\zt \,G(\xt-\zt)\rho_A^a(\zt,z^-) t^a\right].\label{wl}
\end{align}
Here $G(\xt-\zt)$ is the Green’s function for the 2-dimensional Laplace operator, which is inverted in order to relate $\rho_A$ to the Wilson lines:
\begin{align}
G(\xt-\yt)=\int\frac{d^2\kt}{(2\pi)^2}\frac{\exp [i\kt\cdot(\xt-\yt)]}{\kt^2+m^2}.
\end{align}
In doing this, it is necessary to suppress the Coulomb-like tails of the resulting color fields. This is done by introducing an IR regulator $m$, which can be interpreted as a confinement scale of order $\Lambda_{\text{QCD}}$.

Along with the color charge densities, Wilson lines constitute the basic elements of our theoretical framework. More specifically, another building block of our calculations is the two-point function of Wilson lines, known as dipole amplitude:
\begin{align}
D(\xt-\yt)=\frac{1}{N_c}\left\langle U^{\dagger}(\xt)U(\yt) \right\rangle.
\end{align}
There are several parametrizations available for this object in the literature. Although it is possible to compute it explicitly within the MV model, in the case of the nucleus we adopt the GBW distribution \cite{Golec-Biernat:1998zce}:
\begin{align}
D_{\text{GBW}}(\xt-\yt)=\exp\left[-\frac{Q_s^2(\xt-\yt)^2}{4}\right],\label{GBWdip}
\end{align}
as it gives rise to simpler final expressions. By doing this, we effectively set the saturation scale $Q_s$ as the only scale characterizing the nucleus, thus leaving $\mu^2$ and $m$ as proton-exclusive parameters in the remaining sections.

\subsection{Proton: Hot spots + MV model}

Basically the same kinematic considerations as outlined above apply to the case of the proton, up to a $x^-\!\rightarrow\!x^+$ change (due to the proton moving in the opposite direction than the nucleus). Also, we denote the light cone gauge field carried by the proton as:
\begin{align}
\beta^{i}(\xt)=-\frac{1}{ig}V(\xt)\partial^{i}V^{\dagger}(\xt),\label{gfieldp}
\end{align}
with
\begin{align}
V(\xt)={\cal P}^+\exp\left[ -ig\int^{\infty}_{-\infty}dz^+\int d^2\zt \,G(\xt-\zt)\rho^a(\zt,z^+) t^a\right]\label{wlp}
\end{align}
being the corresponding Wilson line. Apart from this change of notation, the difference between our descriptions of proton and nucleus lies fundamentally in their transverse structures, encoded in $\rho^a$ and $\rho_A^a$ respectively. Whereas $\rho_A^a$ is modeled as a dense and infinitely large homogenous distribution, $\rho^a$ is given an inhomogeneous, nontrivial transverse structure that can be treated as dilute on average.

We will use the hot spot model formulated in Ref.\ \cite{Demirci:2021kya} to give the proton a nontrivial transverse structure. This model allows to incorporate hot spot position fluctuations into one's computation. The physically relevant observables are then found by performing a double averaging procedure, where one performs the CGC average discussed above and subsequently the hot spot average. This model allows the computation to be performed with any positive number of hot spots $(N_q)$, but in this work we are mainly focusing on $N_q=3$, as this is the number of valence quarks (which are the inspiration for this model). These hot spots are argued to be carrying the point-like color charges of the proton.

Using the CGC description, the proton is divided into the color charge-carrying Gaussian hot spots and the gluon fields produced by these color charges. The hot spots enter the calculation through the color charge density two-point correlators as:
\begin{equation} \label{eq:HSRhoTwoPoint}
\left\langle \rho^a \left( \xt \right) \rho^b \left( \yt \right) \right\rangle_{CGC}
=
\sum_{i=1}^{N_q} \mu^2 \left( \frac{\xt+\yt}{2}-\bt_i \right)
\delta^{ \left( 2 \right)} \left( \xt-\yt \right)\delta^{ab}.
\end{equation}
Here $\mu^2$ describes the profile of the Gaussian hot spot, $N_q$ is the number of hot spots and $\bt_i$ denote the coordinates of the centers of the hot spots. This correlator is basically the same as in the MV model, but with the added transverse profile for the color charges of the proton. The exact profile of the hot spot reads:
\begin{equation} \label{eq:HotSpotGaussian}
\mu^2(\xt) = \frac{\mu^2_0}{2\pi r^2} \exp \left[ -\frac{\xt^2}{2r^2} \right].
\end{equation}
Here $r$ denotes the radius of the hot spot and $\mu^2_0$ is a parameter characterizing the amount of color charge in it. Note that the dimensions of $\mu^2$ and $\mu^2_0$ are different.

The fluctuations of the positions of the hot spots are also assumed to obey Gaussian statistics. We fix the center of the proton $(\Bt)$ to be at the center of mass of the hot spots and let them be distributed Gaussianly around it, only constrained by the center of mass requirement. The Gaussian distribution of the hot spot positions is characterized by the weight
\begin{equation} \label{eq:HSDistrWeight}
T(\xt) = \frac{1}{2\pi R^2}\exp \left[ -\frac{\xt^2}{2R^2} \right],
\end{equation}
where $R$ is the proton size parameter dictating how far from the center of the proton the hot spots can reside.

Because the positions of the hot spots fluctuate from event to event, we need to average over their fluctuations to get physically relevant observables. For this end we define the following double average:
\begin{equation} \label{eq:DoubleAverage}
\left\langle \left\langle \mathcal{O} \right\rangle \right\rangle \equiv \left\langle \left\langle \mathcal{O} \right\rangle_{\text{CGC}} \right\rangle_{\text{Hotspot}}  = \left( \frac{2\pi R^2}{N_q} \right) \int \prod_{i=1}^{N_q} \left[ d^2\bt_i T \left( \bt_i - \Bt \right) \right] \delta^{\left( 2 \right)} \left( \frac{1}{N_q}\sum^{N_q}_{j=1} \bt_j - \Bt \right) \left\langle \mathcal{O} \right\rangle_{\text{CGC}}.
\end{equation}
Here the prefactor is chosen to ensure that $\left\langle \left\langle 1 \right\rangle \right\rangle = 1$. In this definition we can see explicitly how we let each of the hot spots fluctuate Gaussianly around the center $\Bt$, with the restriction that the center of mass of the hot spot system is fixed at the center of the proton $\Bt$. Therefore, the procedure to calculate an observable consists in: first, computing its CGC average using the MV model two-point function with the additional input of the transverse structure (\eqref{eq:HSRhoTwoPoint}); and then, averaging over the hot spot locations.

\subsection{Freely-evolving Glasma}
Despite the relative simplicity of our treatment of proton and nucleus, the Glasma state that results from their collision is a complex, strongly out-of-equilibrium medium whose $\tau$-evolution cannot be computed analytically without adopting some approximations. In this section we discuss the ones we applied.

In the CGC, the dynamics of the fields resulting from a pA collision are described by the CYM equations with two sources:
\begin{align}
[D_{\mu},F^{\mu\nu}]=J^{\nu}_A+J^{\nu}_p\approx \delta^{+\nu}\rho_A^a(\xt)\delta(x^-)t^a+\delta^{-\nu}\rho^a(\xt)\delta(x^+)t^a.\label{ym01}
\end{align}
Although these equations do not have a general analytical solution, it is possible to obtain exact expressions for the Glasma fields at $\tau\!=\!0^+$ by matching the $\tau\!\rightarrow\!0$ singularities in the left and right hand sides of \eqref{ym01} (see e.g.\ \cite{Kovner:1995ts,Kovner:1995ja} for details).
By doing this in the Fock-Schwinger gauge $A^\tau\!=\!0$ (which works as an interpolation of the light cone gauges of proton and nucleus), one obtains:
\begin{align}
\label{eq:kmw1}
A^{i} (\tau=0^+\!,\xt)&\equiv A^i_0(\xt)= \alpha^{i}(\xt)+\beta^{i}(\xt)\\
\label{eq:kmw2}
A (\tau=0^+\!,\xt)&\equiv A_0(\xt)= \frac{ig}{2} \left[ \alpha^{i}(\xt) , \beta^{i}(\xt) \right].
\end{align}
A notable aspect about this solution is that it gives rise to purely longitudinal electric and magnetic fields:
\begin{align}
E^{\eta}(\tau=0^+,\xt)\!\equiv\!E^{\eta}_0(\xt)\!=&-ig\,\delta^{ij}[\alpha^{i}(\xt),\beta^{j}(\xt)]\label{elmag1}\\
B^{\eta}(\tau=0^+,\xt)\!\equiv\!B^{\eta}_0(\xt)\!=&-ig\,\epsilon^{ij}[\alpha^{i}(\xt),\beta^{j}(\xt)],\label{elmag2}
\end{align}
which are the only nonzero components of the Glasma strength tensor at $\tau\!=\!0^+$.
The above expressions serve as boundary conditions for the evolution in the forward light-cone, $\tau\!\geq\!0^+$. In this region the Yang-Mills equations become homogeneous owing to the longitudinal support of our charge densities $\rho$, $\rho_A$ being Lorentz-contracted down to delta functions. Thus, in the comoving coordinate system $(\tau,\eta,i)$ the separate components of \eqref{ym01} read:
\begin{align}
ig\tau\left[ A, \partial_{\tau}A\right]-\frac{1}{\tau}\left[ D_{i},\partial_{\tau}A^{i} \right]&=0\label{comov3a}\\
\frac{1}{\tau}  \partial_{\tau}\frac{1}{\tau}\partial_{\tau}(\tau^2 A) -\left[ D_{i},\left[ D^{i},A \right]\right]&=0\label{comov3b}\\
\frac{1}{\tau}  \partial_{\tau}(\tau \partial_{\tau}A^{i}) -ig\tau^2\left[ A,\left[ D^{i},A \right]\right]-\left[ D^{j},F^{j i}\right]&=0.\label{comov3}
\end{align}
In order to find an analytical solution to this system we will neglect all terms of higher order in the Glasma potentials, which necessarily introduces a gauge dependence. Let us illustrate this point by transforming to a gauge where Eqs.~(\ref{comov3a}), (\ref{comov3}), (\ref{comov3}) become, respectively:
\begin{align}
\partial_\tau\partial^i\tilde{A}^i&=0\label{eer00}\\
\frac{1}{\tau}  \partial_{\tau}\frac{1}{\tau}\partial_{\tau}(\tau^2 \tilde{A}) -\partial_i\partial^i\tilde{A}&=0\\
\frac{1}{\tau}  \partial_{\tau}(\tau \partial_{\tau}\tilde{A}^{i}) -(\partial^k\partial_k\delta^{ij}-\partial^i\partial^j)\tilde{A}^j&=0.\label{comov4}
\end{align}
For this approximation to be reasonable it is crucial to choose a gauge where the higher orders in $\tilde{A}$, $\tilde{A}^i$ are minimized. Such a choice may be informed by several considerations: for instance, in order to maintain the explicit boost invariance of the Glasma fields, one could consider only $\eta$-independent gauge transformations. One may also exclude $\tau$-dependent transformations for the purpose of preserving the Fock-Schwinger gauge condition. These restrictions narrow the search down to gauges that minimize the transverse components of the fields. Following this line of reasoning, in Ref.\ \cite{Blaizot:2010kh} it was proposed that the transverse Coulomb gauge, defined by the condition $\partial_i\tilde{A}^i\!=\!0$, represents the natural option for our purposes. In this numerical analysis it was found that imposing this gauge choice at $\tau\!=\!0^+$ (in combination with the Fock-Schwinger condition $A^\tau\!=\!0$) minimizes the effect of non-linear dynamics in the forward light-cone, providing a good approximation of the full CYM evolution except for very low momentum modes. This approach is thus well suited for the calculation of the $\tau$-evolution of the Glasma energy density, as it is a largely UV-dominated quantity.

In what follows, we use the notation $\tilde{A}$ to denote the Glasma fields in the Coulomb gauge, leaving the explicit transformation to be specified later. We can now rewrite the Yang-Mills equations as:
\begin{align}
\frac{1}{\tau}  \partial_{\tau}\frac{1}{\tau}\partial_{\tau}(\tau^2 \tilde{A}) -\partial_{i}\partial^{i}\tilde{A}&=0\label{eer1}\\
\frac{1}{\tau}  \partial_{\tau}(\tau \partial_{\tau}\tilde{A}^{i}) -(\partial^{k}\partial_{k}\delta^{ij}-\partial^{i}\partial^{j})\tilde{A}^{j}&=0.\label{eer2}
\end{align}
This system can be solved in momentum space as:
\begin{align}
\tilde{A}(\tau,\kt) &=\tilde{A}_0(\kt) \frac{2 J_1(k\tau)}{k\tau}\label{sollsS_v11}\\
\tilde{A}^i(\tau,\kt) &=\tilde{A}^i_0(\kt)J_0(k\tau),\label{sollsS_v2}
\end{align}
where we assumed the Glasma fields to propagate as free plane waves.
The initial conditions $\tilde{A}_0$, $\tilde{A}_0^i$ can then be computed by matching these fields to the $\tau\!=\!0^+$ fields in momentum space. As a result, the following $\tau>0$ solution is obtained:
\begin{align}
\tilde{A}(\tau,\kt) &=\frac{\tau}{k}\tilde{E}^{\eta}_0(\kt)J_1(k\tau)\\
\tilde{A}^i(\tau,\kt) &=-i\frac{\epsilon^{ij}\kt^j}{k^2}\tilde{B}^{\eta}_0(\kt)J_0(k\tau),\label{solls2}
\end{align}
where $\tilde{E}^{\eta}_0$, $\tilde{B}^{\eta}_0$ correspond to Eqs.\ (\ref{elmag1}), (\ref{elmag2}) transformed to Coulomb gauge. From here one can calculate the full electric and magnetic fields as functions of $\tau$: 
\begin{align}
E^{\eta}(\tau,\kt)=&\,E^{\eta}_0(\kt)J_0(k\tau)\\
E^{i}(\tau,\kt)=&\,-i\epsilon^{ij}\frac{\kt^{j}}{k}B^{\eta}_0(\kt)J_1(k\tau)\\
B^{\eta}(\tau,\kt)=&\,B^{\eta}_0(\kt)J_0(k\tau)\\
B^{i}(\tau,\kt)=&\,-i\epsilon^{ij}\frac{\kt^{j}}{k}E^{\eta}_0(\kt)J_1(k\tau).
\end{align}
In order to work in coordinate space we perform a Fourier transform of the initial conditions. For example, for $E^{\eta}_0$ we have:
\begin{align}
E^{\eta}_0(\xt)\!&=\!-ig\delta^{ij}\!\!\int \!\frac{d^2\kt}{(2\pi)^2}\!\!\int \!d^2\ut\,[\alpha^{i}(\ut),\beta^{j}(\ut)]e^{i\kt\cdot(\xt-\ut)}\equiv\!\int \frac{d^2\kt}{(2\pi)^2}E^{\eta}_0(\kt)e^{i\kt\cdot\xt},
\end{align}
and thus, we get to:
\begin{align}
E^{\eta}(\tau,\xt)\!=\!\!\int \!\frac{d^2\kt}{(2\pi)^2}\!\!\int &\!d^2\ut\,E^{\eta}_0(\ut)J_0(k\tau)e^{i\kt\cdot(\xt-\ut)}.
\end{align}
Similar expressions are obtained for $E^{i}$ and the magnetic fields. At this point it is necessary to explicitly transform our initial conditions (Eqs.\ (\ref{elmag1}), (\ref{elmag2})) to Coulomb gauge. Note that, in their linearized version, the equations of motion explicitly conserve the Coulomb gauge condition (\eqref{eer00}) and therefore, it is irrelevant at which value of $\tau$ we impose it. As previously done in \cite{Lappi:2017skr}, we do this at $\tau\!=\!0^+$ by considering a general gauge transformation driven by $V^{\dagger}U^{\dagger}$:
\begin{align}
\tilde{E}^{\eta}(\tau=0^+,\xt)\!=\!-ig\delta^{ij}V^{\dagger}U^{\dagger}\left[\alpha^{i},\beta^{j}\right]VU\\
\tilde{B}^{\eta}(\tau=0^+,\xt)\!=\!-ig\epsilon^{ij}V^{\dagger}U^{\dagger}\left[\alpha^{i},\beta^{j}\right]VU.
\end{align}
By considering the proton to be dilute, we can expand $V$ to the lowest order in the sources, obtaining:
\begin{align}
\tilde{E}^{\eta}(\tau=0^+,\xt)\!\approx\!-ig\delta^{ij}U^{\dagger}\left[\alpha^{i},\beta^{j}\right]U=\delta^{ij}\beta^{j,a}(\xt)\partial^{i}U^{ab}(\xt)t^b\\
\tilde{B}^{\eta}(\tau=0^+,\xt)\!\approx\!-ig\epsilon^{ij}U^{\dagger}\left[\alpha^{i},\beta^{j}\right]U=\epsilon^{ij}\beta^{j,a}(\xt)\partial^{i}U^{ab}(\xt)t^b.
\end{align}
In order to obtain the last equality we have substituted \eqref{gfield0} and used the identity $U^{\dagger}t^{a}U\!=\!U^{ab}t^{b}$.

\section{Time evolution of energy density correlators}\label{tau_ev}
Let us now apply the formalism introduced above to the calculation of $\tau$-dependent correlators of the energy density of the system, defined as:
\begin{align}
\varepsilon(\tau,\xt)=\text{Tr}\{E^{\eta}E^{\eta}\!+\!B^{\eta}B^{\eta}\!+\!E^iE^i\!+\!B^iB^i\}.
\end{align}
\subsection{One-point function}
We start with the average value of $\varepsilon$:
\begin{align}
\langle\varepsilon(\tau,\xt)\rangle\!=&\,\frac{1}{2}(\delta^{ij}\delta^{kl}\!+\epsilon^{ij}\epsilon^{kl})\!\!\!\int\limits_{\pt,\kt}\int\limits_{\ut,\vt}\langle\partial^iU^{ab}(\ut)\partial^kU^{cb}(\vt)\rangle\,\langle\beta^{j,a}(\ut)\beta^{l,c}(\vt)\rangle\nonumber\\
&\times\left(J_0(p\tau)J_0(k\tau)-\frac{\pt\!\cdot\!\kt}{p\,k}J_1(p\tau)J_1(k\tau)\right)e^{i\pt\cdot(\xt-\ut)}e^{i\kt\cdot(\xt-\vt)}.\label{tauEd}
\end{align}
Here $\int_{\pt}$ corresponds to the integration over momentum $\int\frac{d^2\pt}{(2\pi)^2}$, while $\int_{\ut}$ stands for $\int d^2\ut$.  This expression features Fourier transforms of products of Bessel functions of zeroth and first order, which can be calculated analytically. Integrating over the angular variables $\theta_{\pt}$ and $\theta_{\kt}$, the second line of \eqref{tauEd} becomes:
\begin{align}
\int\frac{dp\,p}{(2\pi)}\frac{dk\,k}{(2\pi)}\bigg(J_0(|\xt-\ut|p)J_0(|\xt-\vt|k)J_0(p\tau)J_0(k\tau)+\cos{(\theta_{\xt-\ut}\!-\theta_{\xt-\vt})}J_1(|\xt\!-\!\ut|p)J_1(|\xt\!-\!\vt|k)J_1(p\tau)J_1(k\tau)\!\bigg).
\end{align}
Then, we can apply the orthogonality condition of the Bessel functions:
\begin{align}
\int^{\infty}_{0}J_{\nu}(kr)J_{\nu}(sr)rdr=\frac{\delta(k-s)}{s},
\end{align}
obtaining:
\begin{align}
\hspace{-0.2cm}\langle\varepsilon(\tau,\xt)\rangle\!=\!\frac{1}{2}(\delta^{ij}\delta^{kl}\!+\epsilon^{ij}\epsilon^{kl})\!\!\int\limits_{\ut,\vt}\!\!\langle\partial^iU^{ab}(\ut)\partial^kU^{cb}(\vt)\rangle\,\langle\beta^{j,a}(\ut)\beta^{l,c}(\vt)\rangle\nonumber\\
\times\!\frac{\delta(|\xt-\ut|\!-\!\tau)}{2\pi\tau}\frac{\delta(|\xt-\vt|\!-\!\tau)}{2\pi\tau}(1+\cos(\theta_{\xt-\ut}\!-\theta_{\xt-\vt})).\label{tauEd1}
\end{align}
In order to integrate out the Dirac deltas we perform a change of variables from ($\ut$, $\vt$) to ($\st\!=\!\xt-\ut$, $\ttt\!=\xt-\vt$). In a collision between two infinite-sized objects, this process would explicitly remove the dependence on the specific transverse point $\xt$, as the correlators describing each side of the interaction would then depend only on the relative distance $\ut-\vt$. However, that does not necessarily happen when considering a collision where one of the objects has a finite size. That is the case of the proton, represented here by the two-point function $\langle\beta^{j,a}(\ut)\beta^{l,c}(\vt)\rangle$. After the variable change, this object carries the dependence on the point $\xt$:
\begin{align}
\hspace{-0.2cm}\langle\varepsilon(\tau,\xt)\rangle\!=\!\frac{1}{2}(\delta^{ij}\delta^{kl}\!+\epsilon^{ij}\epsilon^{kl})\!\!\int\limits_{\st,\ttt}\!\!\langle\partial^iU^{ab}(\st)\partial^kU^{cb}(\ttt)\rangle\,\langle\beta^{j,a}(\xt-\st)\beta^{l,c}(\xt-\ttt)\rangle\nonumber\\
\times\!\frac{\delta(s\!-\!\tau)}{2\pi\tau}\frac{\delta(t\!-\!\tau)}{2\pi\tau}(1+\cos(\theta_{\st}\!-\theta_{\ttt})).\label{tauEd2}
\end{align}
From the above expressions we can see how $\langle\varepsilon\rangle$ results from the interference of two correlators that represent the gluon content of proton and nucleus. In the case of the nucleus (and in the dilute-dense limit we have adopted), this correlator corresponds to the adjoint dipole distribution affected by the partial derivatives $\partial^{i}_{\st}\partial^{j}_{\ttt}$. By adopting the GBW model, we can compute this object as:
\begin{align}
\partial^i_{\st}\partial^k_{\ttt}\langle U^{ab}(\st)U^{cb}(\ttt)\rangle=\frac{\delta^{ac}}{2}\left(\delta^{ik}Q^2_{s,\text{A}}-\frac{Q^4_{s,\text{A}}}{2}(\st-\ttt)^{i}(\st-\ttt)^{k}\right)\exp\left[-\frac{Q^2_{s,\text{A}}}{4}|\st-\ttt|^2\right].\label{Udipol}
\end{align}
Here we have introduced the GBW distribution corresponding to adjoint Wilson lines. Note that, with our notation, the color factor that would set \eqref{GBWdip} apart from its adjoint version is absorbed in a re-definition of $Q_s$. Thus, the adjoint saturation scale $Q_{s,\text{A}}$ is related to the fundamental one by $Q^2_{s,\text{A}}\!=\!\frac{C_A}{C_F}Q^2_{s,\text{F}}=\frac{2N^2_c}{N^2_c-1}Q^2_{s,\text{F}}$. As in our calculations we will only need $Q^2_{s,\text{A}}$, in the following we will denote it simply by $Q_s$ in order to alleviate our notation.

Let us now focus on the dilute side contribution, which we compute by considering a hot spot model. As explained in \secref{th_f}, this implies that we promote the simple color charge average to a double average of color charges and hot spot positions:
\begin{align}
\langle\beta^{j,a}(\xt-\st)\beta^{l,c}(\xt-\ttt)\rangle=&\,\left\langle\left\langle \,\,\int\limits_{\zt,\wt}\!\!\partial^{j}_{\st}G(\xt-\st-\zt)\partial^{l}_{\ttt}G(\xt-\ttt-\wt)\rho^{a}(\zt)\rho^{c}(\wt) \right\rangle\right\rangle\nonumber\\
=&\,\delta^{ac}N_q\int\limits_{\zt}\!\!\partial^{j}_{\st}G(\xt-\st-\zt)\partial^{l}_{\ttt}G(\xt-\ttt-\zt)F_1(\zt,\Bt).\label{proton1}
\end{align}
Here, using the same notation as Ref.\cite{Demirci:2021kya}, we introduce the function $F_1(\zt,\Bt)$, defined as:
\begin{align}\label{funcf1}
F_1(\zt,\Bt)\equiv\langle\mu^2(\zt-\bt_i)\rangle_{\text{Hotspot}}=\left(\frac{\mu^2_0}{2\pi}\right)\left(\frac{1}{r^2+\frac{N_q-1}{N_q}R^2}\right)\exp\left[-\frac{1}{2}\frac{|\zt-\Bt|^2}{r^2+\frac{N_q-1}{N_q}R^2}\right],
\end{align}
with $i\in\{1,...,N_q\}$.
This function can be interpreted as the average density of the proton color charges given by a single hot spot. As mentioned before, the proton radius $R$, the charge density parameter $\mu_0$, the hot spot size $r$ and the number of hot spots $N_q$ are parameters of the model. The point $\Bt$ is the center of mass of the hot spot system. After shifting the integration variable $\zt$ by $\xt$, we obtain:
\begin{align}
\langle\beta^{j,a}(\st)\beta^{l,c}(\ttt)\rangle=&\,\delta^{ac}N_q\int\limits_{\zt}\!\!\partial^{j}_{\st}G(\st-\zt)\partial^{l}_{\ttt}G(\ttt-\zt)F_1(\zt,\Bt-\xt)\nonumber\\
=&\,\delta^{ac}N_q\left(\frac{\mu^2_0}{2\pi}\right)\left(\frac{1}{r^2+\frac{N_q-1}{N_q}R^2}\right)\int\limits_{\zt}\!\!\partial^{j}_{\st}G(\st-\zt)\partial^{l}_{\ttt}G(\ttt-\zt)\exp\left[-\frac{1}{2}\frac{|\zt-(\Bt-\xt)|^2}{r^2+\frac{N_q-1}{N_q}R^2}\right].\label{proton2}
\end{align}
Substituting Eqs.\ (\ref{proton2}), (\ref{Udipol}) into \eqref{tauEd2} and solving the trivial integrals over $|\st|$, $|\ttt|$, one is left with a couple of integrals over $\theta_{\st}$, $\theta_{\ttt}$, as well as a double integral over $\zt$:
\begin{align} \label{eq:OnePointTauED}
\langle\varepsilon(\tau,\Bt_{\xt})&\rangle\!=\!(N^2_c-1)\int\frac{d\theta_{\st}}{2\pi}\frac{d\theta_{\ttt}}{2\pi}(1+\cos{(\theta_{\st}-\theta_{\ttt})})\left(\frac{Q_s^2}{2}-\frac{Q_s^4\tau^2}{4}(1-\cos{(\theta_{\st}-\theta_{\ttt})})\right)\frac{\tilde{\mu}_0^2}{2\pi}\left(\frac{1}{r^2+\frac{N_q-1}{N_q}R^2}\right)\nonumber\\
&\times\int\limits_{\zt}\!\!\left.\left(\partial^{j}_{\st}G(\st-\zt)\partial^{j}_{\ttt}G(\ttt-\zt)\right)\right|_{s,t=\tau}\exp\left[-\frac{1}{2}\frac{|\zt-\Bt_{\xt}|^2}{r^2+\frac{N_q-1}{N_q}R^2}\right]
\exp\left[-\frac{Q_s^2\tau^2}{2}\left(1-\cos\left(\theta_{\st}-\theta_{\ttt}\right)\right)\right],
\end{align}
where we have defined $\tilde{\mu}_0^2\equiv N_q\mu_0^2$ and $\Bt_{\xt}\equiv \Bt-\xt$.
Now, assuming that $\xt\neq\yt$, we can write:
\begin{align}
\partial^{i}_{\xt}G(\xt-\yt)=-\frac{1}{2\pi}m|\xt-\yt|K_1(m|\xt-\yt|)\frac{(\xt-\yt)^{i}}{|\xt-\yt|^2}\label{res1}\\
\partial^{i}_{\xt}G(\xt-\zt)\partial^{i}_{\yt}G(\yt-\zt)=\left(\frac{m}{2\pi}\right)^2K_1(m|\xt-\zt|)K_1(m|\yt-\zt|)\cos{\left(\theta_{\xt-\zt}-\theta_{\yt-\zt}\right)}\label{res1bb},
\end{align}
where $K_1$ is the modified Bessel function of the second kind and order 1. Note that \eqref{res1} diverges in the UV limit. We regularized this behavior through a short-distance cut-off $C_0$, introduced by (implicitly) multiplying by the Heaviside step function $\Theta(|\xt-\yt|\!-\!C_0)$. Substituting \eqref{res1bb} in \eqref{eq:OnePointTauED}, we get to the final result of this section:
\begin{align}
\langle\varepsilon(\tau,&\,\Bt_{\xt})\rangle=\!(N^2_c-1)\int\frac{d\theta_{\st}}{2\pi}\frac{d\theta_{\ttt}}{2\pi}(1+\cos{(\theta_{\st}-\theta_{\ttt})})\left(\frac{Q_s^2}{2}-\frac{Q_s^4\tau^2}{4}(1-\cos{(\theta_{\st}-\theta_{\ttt})})\right)\frac{\tilde{\mu}_0^2}{2\pi}\left(\frac{1}{r^2+\frac{N_q-1}{N_q}R^2}\right)\nonumber\\
&\times\!\!\int\limits_{\zt}\!\left(\frac{m}{2\pi}\right)^2\!\!K_1(m|\tau\sthat-\zt|)K_1(m|\tau\tthat-\zt|)\cos{\left(\theta_{\st-\zt}-\theta_{\ttt-\zt}\right)}\exp\!\left[-\frac{1}{2}\frac{|\zt-\Bt_{\xt}|^2}{r^2+\frac{N_q-1}{N_q}R^2}\right]\!\exp\!\left[-\frac{Q_s^2\tau^2}{2}\left(1\!-\!\cos\left(\theta_{\st}\!-\!\theta_{\ttt}\right)\right)\right]\!.
\end{align}
Note that the previous correlator depends only on the proper time $\tau$ and the relative distance between $\xt$ and the center of mass of the proton, $\Bt_{\xt}$. If we take $\tau=0$, one can explicitly perform the integrals over $\theta_{s,t,z}$, obtaining:
\begin{align}
\langle\varepsilon(\tau=0,B_{x})\rangle&\,=\!\frac{Q_s^2}{2}\tilde{\mu}_0^2\frac{(N^2_c-1)}{r^2+\frac{N_q-1}{N_q}R^2}\int dz\,z\left(\frac{m}{2\pi}\right)^2\!\!(K_1(mz))^2I_0\!\left(\frac{z\,B_x}{r^2+\frac{N_q-1}{N_q}R^2}\right)\!\!\exp\!\left[-\frac{1}{2}\frac{z^2+B_x^2}{r^2+\frac{N_q-1}{N_q}R^2}\right],
\end{align}
where $I_0$ is the modified Bessel function of the first kind and order 0.

\begin{figure}
    \centering
    \includegraphics[width=0.4\textwidth]{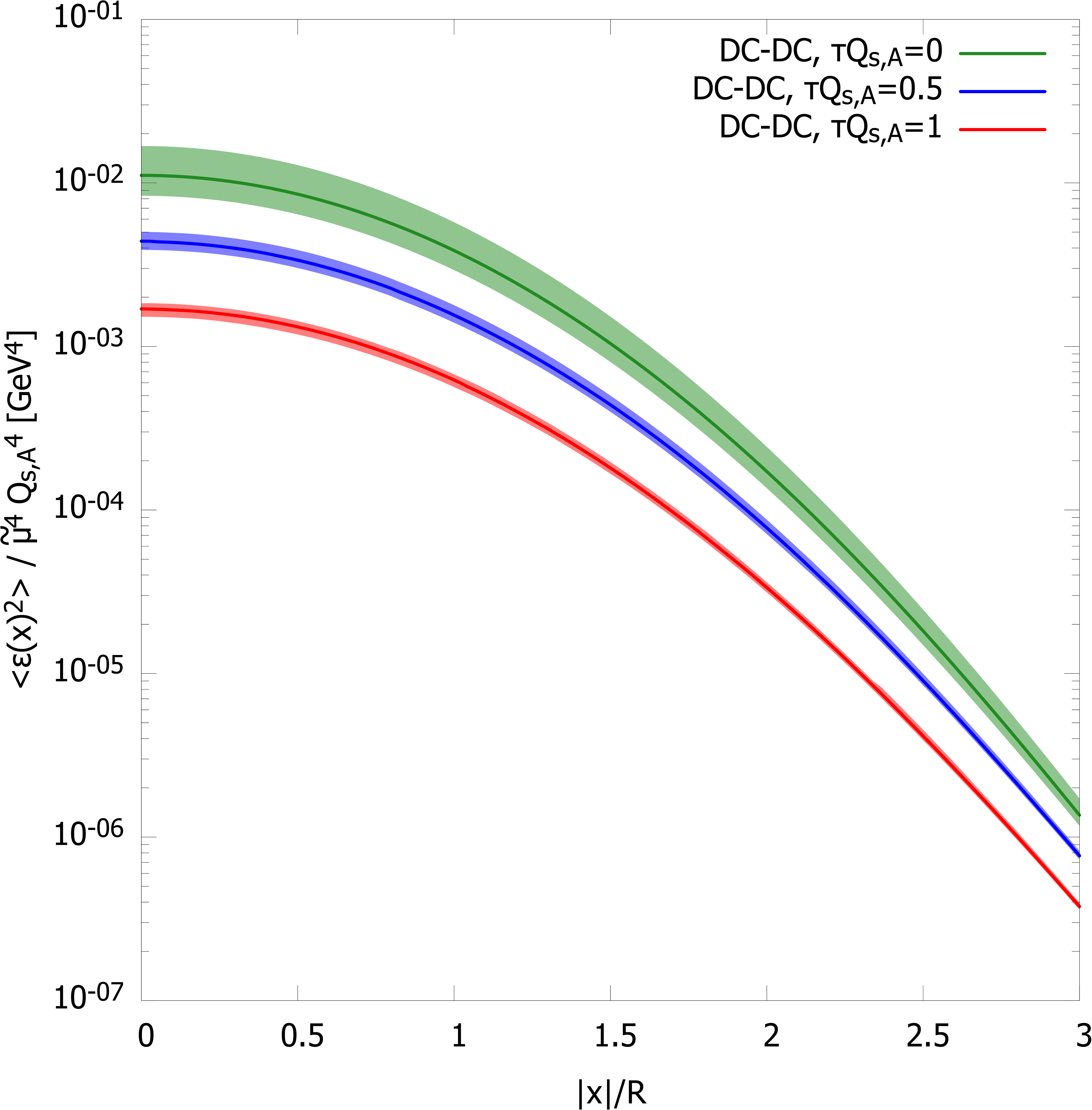}
    \caption{Proper time evolution of average energy density squared as a function of the position relative to the center of mass of the proton. The bands correspond to a variation of $C_0$ by $\pm50\%$. $Q_{s,A}$ stands for the dense nucleus saturation scale.}
    \label{fig:1pf}
\end{figure}

In \figref{fig:1pf} we display the ratio $\langle\varepsilon\rangle^2/(Q_s^2\tilde{\mu}_0^2)^2$ for three values of $\tau$. For this and the remaining plots we will be using the following parameters unless stated otherwise. We have three colors $N_c\!=\!3$ and three hot spots $N_q\!=\!3$. The dense nucleus saturation scale is set to be $Q_s\!=\!3 \text{ GeV}$, the IR regulator is $m\!=\!0.5\text{ GeV}$ and the UV regulator is $C_0\!=\!0.05 \text{ GeV}^{-1}$. The proton radius parameter is $R\!=\!\sqrt{3.3}\text{ GeV}^{-1}$ and the hot spot radius is $r\!=\!\sqrt{0.7}\text{ GeV}^{-1}$. The proton and hot spot size parameters were originally taken from \cite{Mantysaari:2016jaz}.

\subsection{Two-point function}\label{2pfsec}

Let us now compute the energy density two-point function:
\begin{align}
\langle\varepsilon(\tau,\xt)\varepsilon(\tau,\yt)\rangle\!=&\,\frac{1}{4}(\delta^{ij}\delta^{kl}\!+\epsilon^{ij}\epsilon^{kl})(\delta^{i'\!j'\!}\delta^{k'\!l'\!}\!+\epsilon^{i'\!j'\!}\epsilon^{k'\!l'\!})\!\!\!\int\limits_{\pt,\kt}\int\limits_{\ptbar,\ktbar}\int\limits_{\ut,\vt}\int\limits_{\utbar,\vtbar}\langle\partial^iU^{ab}_{\ut}\partial^kU^{cb}_{\utbar}\partial^{i'}U^{a'b'}_{\vt}\partial^{k'}U^{c'b'}_{\vtbar}\rangle\langle\beta^{j,a}_{\ut}\beta^{l,c}_{\utbar}\beta^{j'\!\!,a'\!\!}_{\vt}\beta^{l'\!\!,c'\!}_{\vtbar}\rangle\nonumber\\
&\times\!\!\left(J_0(p\tau)J_0(\bar{p}\tau)\!-\frac{\pt\!\cdot\!\ptbar}{p\,\bar{p}}J_1(p\tau)J_1(\bar{p}\tau)\right)\!\!\left(J_0(k\tau)J_0(\bar{k}\tau)\!-\frac{\kt\!\cdot\!\ktbar}{k\,\bar{k}}J_1(k\tau)J_1(\bar{k}\tau)\right)\nonumber\\
&\times\!e^{i\pt\cdot(\xt-\ut)}e^{i\kt\cdot(\yt-\vt)}e^{i\ptbar\cdot(\xt-\utbar)}e^{i\ktbar\cdot(\yt-\vtbar)}\nonumber\\
=&\,\frac{1}{4}(\delta^{ij}\delta^{kl}\!+\epsilon^{ij}\epsilon^{kl})(\delta^{i'\!j'\!}\delta^{k'\!l'\!}\!+\epsilon^{i'\!j'\!}\epsilon^{k'\!l'\!})\!\!\int\limits_{\ut,\vt}\int\limits_{\utbar,\vtbar}\langle\partial^iU^{ab}_{\ut}\partial^kU^{cb}_{\utbar}\partial^{i'}U^{a'b'}_{\vt}\partial^{k'}U^{c'b'}_{\vtbar}\rangle\langle\beta^{j,a}_{\ut}\beta^{l,c}_{\utbar}\beta^{j'\!\!,a'\!\!}_{\vt}\beta^{l'\!\!,c'\!}_{\vtbar}\rangle\nonumber\\
&\times\!\frac{\delta(|\xt-\ut|\!-\!\tau)}{2\pi\tau}\frac{\delta(|\xt-\utbar|\!-\!\tau)}{2\pi\tau}\frac{\delta(|\yt-\vt|\!-\!\tau)}{2\pi\tau}\frac{\delta(|\yt-\vtbar|\!-\!\tau)}{2\pi\tau}\nonumber\\
&\times(1+\cos(\theta_{\xt-\ut}-\theta_{\xt-\utbar}))(1+\cos(\theta_{\yt-\vt}-\theta_{\yt-\vtbar})),\label{tauEd2p}
\end{align}
where we have adopted the shorthand notation $\beta^{i,a}_{\xt}\!\equiv\!\beta^{i,a}(\xt)$ (and ditto for the Wilson lines). As we did in the case of the one-point function, we perform a variable change from ($\ut$, $\utbar$, $\vt$, $\vtbar$) to ($\st\!=\!\xt-\ut$, $\stbar\!=\!\xt-\utbar$, $\ttt\!=\yt-\vt$, $\ttbar\!=\yt-\vtbar$), which will allow us to straightforwardly integrate over the Dirac deltas:
\begin{align}
\langle\varepsilon(\tau,\xt)\varepsilon(\tau,\yt)\rangle\!=&\,\frac{1}{4}(\delta^{ij}\delta^{kl}\!+\epsilon^{ij}\epsilon^{kl})(\delta^{i'\!j'\!}\delta^{k'\!l'\!}\!+\epsilon^{i'\!j'\!}\epsilon^{k'\!l'\!})\!\!\int\limits_{\st,\ttt}\int\limits_{\stbar,\ttbar}\langle\partial^iU^{ab}_{\xt-\st}\partial^kU^{cb}_{\xt-\stbar}\partial^{i'}U^{a'b'}_{\yt-\ttt}\partial^{k'}U^{c'b'}_{\yt-\ttbar}\rangle\langle\beta^{j,a}_{\xt-\st}\beta^{l,c}_{\xt-\stbar}\beta^{j'\!\!,a'\!\!}_{\yt-\ttt}\beta^{l'\!\!,c'\!}_{\yt-\ttbar}\rangle\nonumber\\
&\times\!\frac{\delta(s\!-\!\tau)}{2\pi\tau}\frac{\delta(\bar{s}\!-\!\tau)}{2\pi\tau}\frac{\delta(t\!-\!\tau)}{2\pi\tau}\frac{\delta(\bar{t}\!-\!\tau)}{2\pi\tau}(1+\cos(\theta_{\st}-\theta_{\stbar}))(1+\cos(\theta_{\ttt}-\theta_{\ttbar})).\label{tauEd2p2}
\end{align}
We will separate the proton and the nucleus-side correlators into their connected and disconnected parts in the same fashion as was done in Ref.\ \cite{Demirci:2021kya} for $\tau\!=\!0^+$. For this we define the following decomposition of the proton-side correlator:
\begin{align} \label{eq:ProtonCorrDecomp}
\langle\beta^{i,a}_{\xt-\st}\beta^{j,b}_{\xt-\stbar}\beta^{k,c}_{\yt-\ttt}\beta^{l,d}_{\yt-\ttbar}\rangle=\langle\beta^{i,a}_{\xt-\st}\beta^{j,b}_{\xt-\stbar}\rangle\langle\beta^{k,c}_{\yt-\ttt}\beta^{l,d}_{\yt-\ttbar}\rangle+\langle\beta^{i,a}_{\xt-\st}\beta^{j,b}_{\xt-\stbar}\beta^{k,c}_{\yt-\ttt}\beta^{l,d}_{\yt-\ttbar}\rangle_{\text{C}},
\end{align}
where the contribution $\langle\beta^{i,a}_{\xt-\st}\beta^{j,b}_{\xt-\stbar}\rangle\langle\beta^{k,c}_{\yt-\ttt}\beta^{l,d}_{\yt-\ttbar}\rangle$ contains the same correlators appearing on the average energy density formula \eqref{tauEd2}. Doing the same for the nucleus-side correlator, we get:
\begin{align} \label{eq:NucleusCorrDecomp}
\langle\partial^iU^{ab}_{\xt-\st}\partial^kU^{cb}_{\xt-\stbar}\partial^{i'}U^{a'b'}_{\yt-\ttt}\partial^{k'}U^{c'b'}_{\yt-\ttbar}\rangle
=
\langle\partial^iU^{ab}_{\xt-\st}\partial^kU^{cb}_{\xt-\stbar} \rangle\langle \partial^{i'}U^{a'b'}_{\yt-\ttt}\partial^{k'}U^{c'b'}_{\yt-\ttbar}\rangle
+
\langle\partial^iU^{ab}_{\xt-\st}\partial^kU^{cb}_{\xt-\stbar}\partial^{i'}U^{a'b'}_{\yt-\ttt}\partial^{k'}U^{c'b'}_{\yt-\ttbar}\rangle_{\text{C}},
\end{align}
where we have the correlators $\langle\partial^iU^{ab}_{\xt-\st}\partial^kU^{cb}_{\xt-\stbar} \rangle\langle \partial^{i'}U^{a'b'}_{\yt-\ttt}\partial^{k'}U^{c'b'}_{\yt-\ttbar}\rangle$, also featured in \eqref{tauEd2}. These parts, composed of products of correlators that appeared previously in the energy density average, are referred to as disconnected contributions. Additionally, the connected parts (denoted by a subscript C) are defined as the full correlator minus the disconnected part.

Now we can decompose the energy density two-point function into a sum of four parts by using \eqref{eq:ProtonCorrDecomp} and \eqref{eq:NucleusCorrDecomp}. We end up with a term where both contributions coming from nucleus and proton sides are disconnected (DC,DC), one where both are connected (C,C), one where the proton side is connected and the nucleus side is disconnected (DC,C) and one where the roles are reversed (C,DC). We call the (DC,C)-part the proton fluctuation part and the (C,DC)-part the nucleus fluctuation part. Note that, with this notation, we have:
\begin{equation}
\langle\varepsilon(\tau,\xt)\varepsilon(\tau,\yt)\rangle _{\text{DC,DC}}
=
\langle \varepsilon(\tau,\xt) \rangle
\langle \varepsilon(\tau,\yt) \rangle .
\end{equation}

We argue that the proton fluctuation part is the dominant source of fluctuations in this model. The (C,C) term is sensitive to both the proton- and the nucleus-side fluctuations and can thus be assumed to give only a small contribution. This is supported by the results obtained in Ref.\ \cite{Demirci:2021kya}, where the same model was used to compute eccentricities in pA collisions at $\tau\!=\!0^+$. It was found that the proton fluctuation contribution is the largest source of eccentricity, with nucleus fluctuations giving only a small correction when the number of hot spots is set to $N_q\!=\!3$. We have no reason to believe that this would not be true at higher values of $\tau$. Thus, we assume that the nucleus-side contribution can be dropped, and that it also keeps the fully connected contribution small. On these grounds, we will only study the eccentricity generated by the hot spot-dominated proton fluctuations.

Let us now examine the proton contribution. The full correlator reads, to lowest order in the dilute proton color sources:
\begin{equation}
\begin{split}
&
\langle \beta_{\xt-\st}^{j,a}\, \beta_{\xt-\stbar}^{l,c}\, \beta_{\yt-\ttt}^{j',a'}\, \beta_{\yt-\ttbar}^{l',c'} \rangle
=
\int d^2\zt d^2\wt \Big[ N_qF_2(\zt,\wt,\Bt)+N_q(N_q-1)F_3(\zt,\wt,\Bt) \Big]
\\ & \times
\Big\{ 
G^{j}_{\st}(\xt-\st-\zt)G^{l}_{\stbar}(\xt-\stbar-\zt)G^{j'}_{\ttt}(\yt-\ttt-\wt)G^{l'}_{\ttbar}(\yt-\ttbar-\wt)\delta^{ac}\delta^{a'c'}
\\ &
+G^{j}_{\st}(\xt-\st-\zt)G^{l}_{\stbar}(\xt-\stbar - \wt)G^{j'}_{\ttt}(\yt-\ttt- \zt)G^{l'}_{\ttbar}(\yt-\ttbar - \wt)\delta^{aa'}\delta^{cc'}
\\ &
+G^{j}_{\st}(\xt-\st-\zt)G^{l}_{\stbar}(\xt-\stbar- \wt)G^{j'}_{\ttt}(\yt-\ttt- \wt)G^{l'}_{\ttbar}(\yt-\ttbar - \zt)\delta^{ac'}\delta^{ca'}
\Big\},
\end{split}
\end{equation}
where $G^{i}_{\st}(\xt-\st)\equiv\partial^{i}_{\st}G(\xt-\st)$.
We perform a variable change from ($\zt$, $\wt$) to ($\ztbar\!=\!\xt\!-\!\zt$, $\wtbar\!=\!\yt\!-\!\wt$), obtaining:
\begin{equation} \label{eq:PConnVarChange}
\begin{split}
&
\langle \beta_{\xt-\st}^{j,a}\, \beta_{\xt-\stbar}^{l,c}\, \beta_{\yt-\ttt}^{j',a'}\, \beta_{\yt-\ttbar}^{l',c'} \rangle
=
\int d^2\bar{\zt} d^2\bar{\wt} \Big[ N_qF_2(\xt-\bar{\zt},\yt-\bar{\wt},\Bt)+N_q(N_q-1)F_3(\xt-\bar{\zt},\yt-\bar{\wt},\Bt) \Big]
\\ & \times
\Big\{ 
G^{j}_{\st}(\bar{\zt}-\st)G^{l}_{\stbar}(\bar{\zt}-\stbar)G^{j'}_{\ttt}(\bar{\wt}-\ttt)G^{l'}_{\ttbar}(\bar{\wt}-\ttbar)\delta^{ac}\delta^{a'c'}
\\ &
+G^{j}_{\st}(\bar{\zt}-\st)G^{l}_{\stbar}((\bar{\wt}-\stbar) +\rt)G^{j'}_{\ttt}((\bar{\zt}-\ttt)- \rt)G^{l'}_{\ttbar}(\bar{\wt}-\ttbar)\delta^{aa'}\delta^{cc'}
\\ &
+G^{j}_{\st}(\bar{\zt}-\st)G^{l}_{\stbar}((\bar{\wt}-\stbar) +\rt)G^{j'}_{\ttt}(\bar{\wt}-\ttt)G^{l'}_{\ttbar}((\bar{\zt}-\ttbar) - \rt)\delta^{ac'}\delta^{ca'}
\Big\},
\end{split}
\end{equation}
with $\rt=\xt-\yt$. Here we distinguish two kinds of contributions, represented by the functions $F_2$, $F_3$:
\begin{align}
F_2(\xt-\bar{\zt},\yt-\bar{\wt},\Bt)\,&=\left(\frac{\mu^2_0}{2\pi r^2}\right)^2\left(\frac{1}{1+2\frac{N_q-1}{N_q}\frac{R^2}{r^2}}\right)\exp\left[-\frac{(2\mathbf{b}-\bar{\zt}-\bar{\wt}-2\Bt)^2}{4r^2\left(1+2\frac{N_q-1}{N_q}\frac{R^2}{r^2}\right)}-\frac{(\rt-(\bar{\zt}-\bar{\wt}))^2}{4r^2}\right]\\
F_3(\xt-\bar{\zt},\yt-\bar{\wt},\Bt)\,&=\left(\frac{\mu^4_0}{(2\pi)^2( R^2+r^2)}\right)\left(\frac{1}{r^2+\frac{N_q-2}{N_q}R^2}\right)\exp\left[-\frac{(2\mathbf{b}-\bar{\zt}-\bar{\wt}-2\Bt)^2}{4\left(r^2+\frac{N_q-2}{N_q}R^2\right)}-\frac{(\rt-(\bar{\zt}-\bar{\wt}))^2}{4(R^2+r^2)}\right],
\end{align}
with $\mathbf{b}=(\xt+\yt)/2$. Analogously to \eqref{funcf1}, these functions are defined as the averages of color charges within the same hot spot ($F_2\equiv\langle\mu^2(\at-\bt_i)\mu^2(\bt-\bt_i)\rangle_{\text{Hotspot}}$) and from two different hot spots ($F_3\equiv\langle\mu^2(\at-\bt_i)\mu^2(\bt-\bt_j)\rangle_{\text{Hotspot}}$). Here, if we take $\xt$, $\yt$ to be opposite to each other on a straight line through the center of the proton, we have $\mathbf{b}=\Bt$ and thus $F_2$, $F_3$ depend only on $\bar{\zt}$, $\bar{\wt}$ and $\rt$.

Having obtained \eqref{eq:PConnVarChange}, we can now compute the proton fluctuation contribution to the energy density two-point function. For this we take the definition of the energy density two-point function \eqref{tauEd2p2} and replace the full dense nucleus correlator with its disconnected contribution and the proton correlator with its connected contribution. Doing this and plugging in the results shown in \eqref{Udipol} and \eqref{eq:PConnVarChange}, we get:
\begin{equation} \label{eq:protonFlucContributionToED2PFunc}
\begin{split}
&
\langle\varepsilon(\tau,\xt)\varepsilon(\tau,\yt)\rangle _{\text{DC,C}}
\\ &
=
\left( N_c^2 -1 \right) \frac{Q_s^4}{4}
\int\limits_{\st,\ttt}
\int\limits_{\bar{\st},\bar{\ttt}}
\int\limits_{\bar{\zt},\bar{\wt}}
\left( 1-\frac{Q_s^2}{4} \left| \st - \bar{\st} \right| ^2 \right)
\left( 1-\frac{Q_s^2}{4} \left| \ttt - \bar{\ttt} \right| ^2 \right)
\\ & \times
\exp \left[ -\frac{Q_s^2}{4} \left( \left| \st - \bar{\st}  \right|^2 + \left| \ttt - \bar{\ttt} \right|^2 \right)\right]
\left(1+\cos(\theta_{\st} - \theta_{\bar{\st}})\right)\left(1+\cos(\theta_{\ttt} - \theta_{\bar{\ttt}})\right)
\\ & \times
\left[
N_q F_2\left( \xt - \bar{\zt}, \yt - \bar{\wt} \right)
+ N_q \left( N_q - 1 \right) F_3 \left( \xt - \bar{\zt}, \yt - \bar{\wt} \right)
\right]
\\ & \times
\Bigg\{ \left( N_c^2 -1 \right)
G^i_{\st} \left( \bar{\zt} - \st \right) 
G^i_{\bar{\st}} \left( \bar{\zt} - \bar{\st} \right)
G^j_{\ttt} \left( \bar{\wt} - \ttt \right) 
G^j_{\bar{\ttt}} \left( \bar{\wt} - \bar{\ttt} \right)
\\ &
+
G^k_{\bar{\st}} \left( \bar{\wt} - \bar{\st} + \rt \right) 
G^k_{\st} \left( \bar{\zt} - \st \right)
\left[
G^l_{\ttt} \left( \bar{\wt} - \ttt \right) 
G^l_{\bar{\ttt}} \left( \bar{\zt} - \bar{\ttt} - \rt \right)
+
G^l_{\bar{\ttt}} \left( \bar{\wt} - \bar{\ttt} \right) 
G^l_{\ttt} \left( \bar{\zt} - \ttt - \rt \right)
\right]
\Bigg\}
\\ & \times
\frac{\delta \left( s - \tau \right)}{2\pi\tau}
\frac{\delta \left( t - \tau \right)}{2\pi\tau}
\frac{\delta \left( \bar{s} - \tau \right)}{2\pi\tau}
\frac{\delta \left( \bar{t} - \tau \right)}{2\pi\tau}
-\langle\varepsilon(\tau,\xt)\varepsilon(\tau,\yt)\rangle _{\text{DC,DC}}.
\end{split}
\end{equation}
This is, by far, the dominant source of fluctuations (and, by extension, of eccentricity) in this model. Here we distinguish two contributions to the proton-side fluctuations: the term proportional to $(N_c^2-1)^2$, which gives us the hot spot fluctuation; and the other two terms in the curly brackets, which give us the color fluctuation. We note that the structure of the Green's functions product in the hot spot contribution is similar to that of the energy density one-point function \eqref{eq:OnePointTauED}, and thus we argue that the color charge fluctuations are averaged over in this term.

\begin{figure}
\centering
\includegraphics[width=0.45\textwidth]{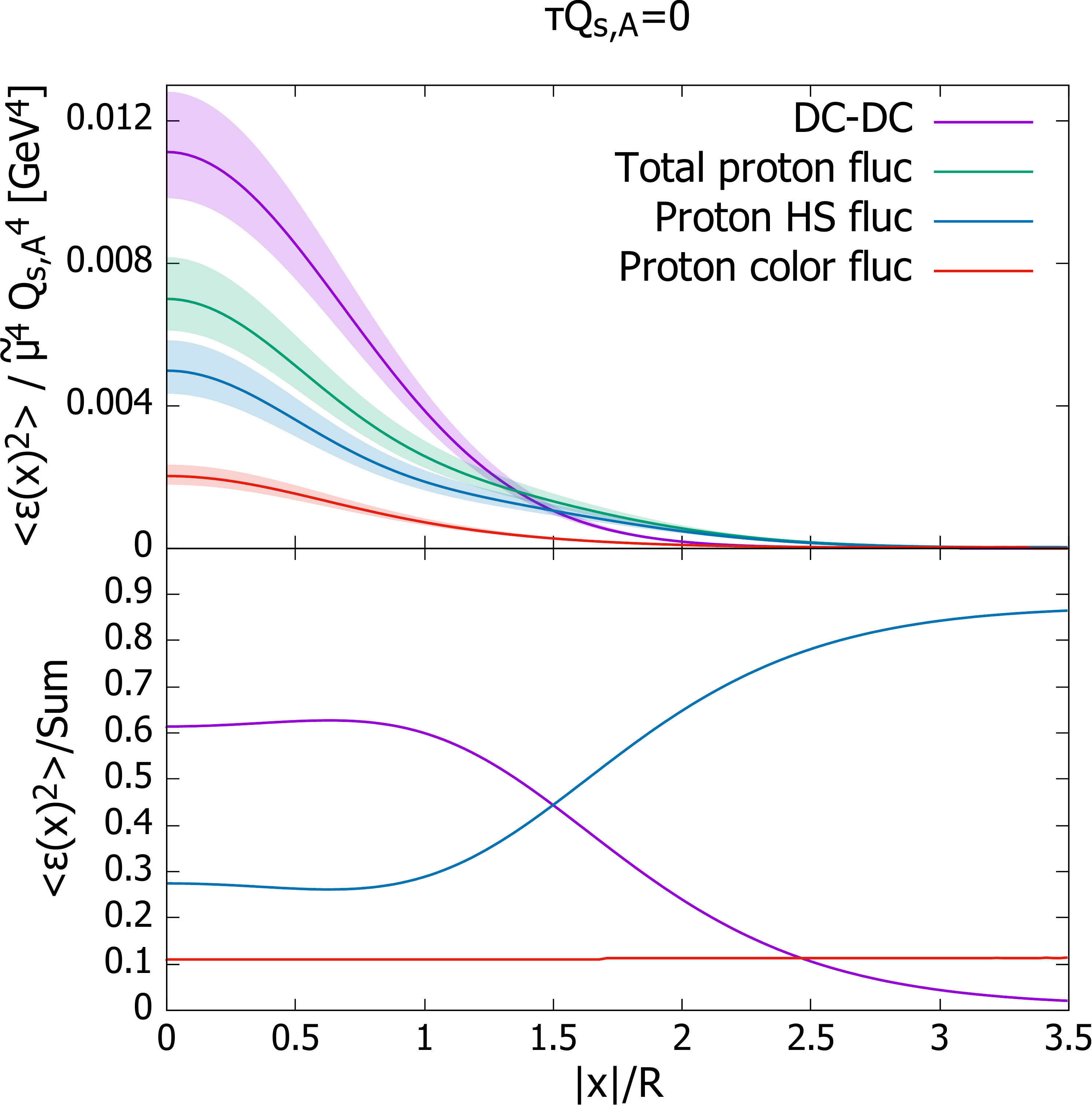}
\includegraphics[width=0.45\textwidth]{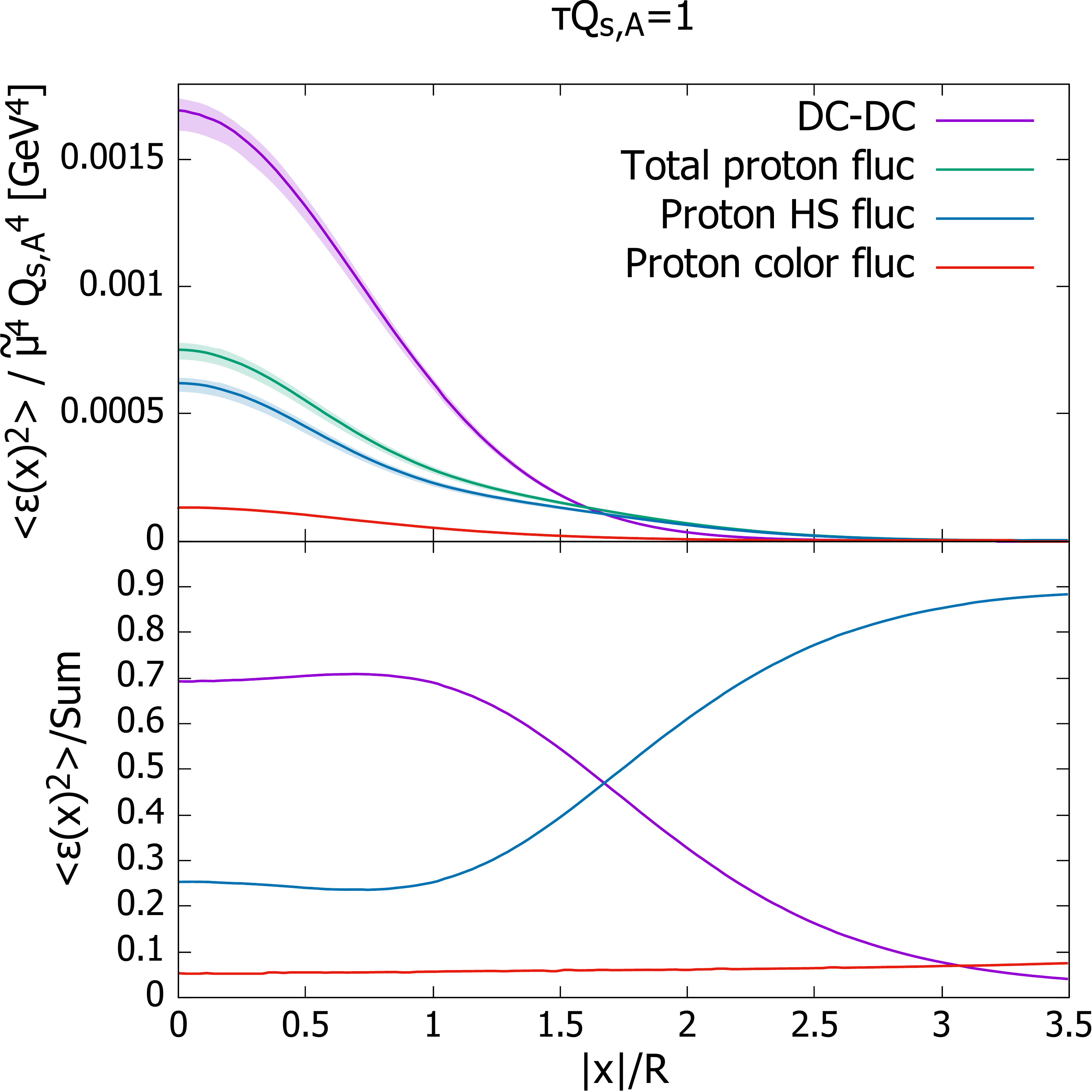} \\
\caption{Local energy density two-point function with proton fluctuation contribution separated into hot spot and color fluctuation parts. The total proton fluctuation and the fully disconnected contribution is also shown for reference. The bands correspond to a variation of $C_0$ by $\pm20\%$.}
\label{fig:2PointHSAndColor}
\end{figure}

\begin{figure}
\centering
\includegraphics[width=0.45\textwidth]{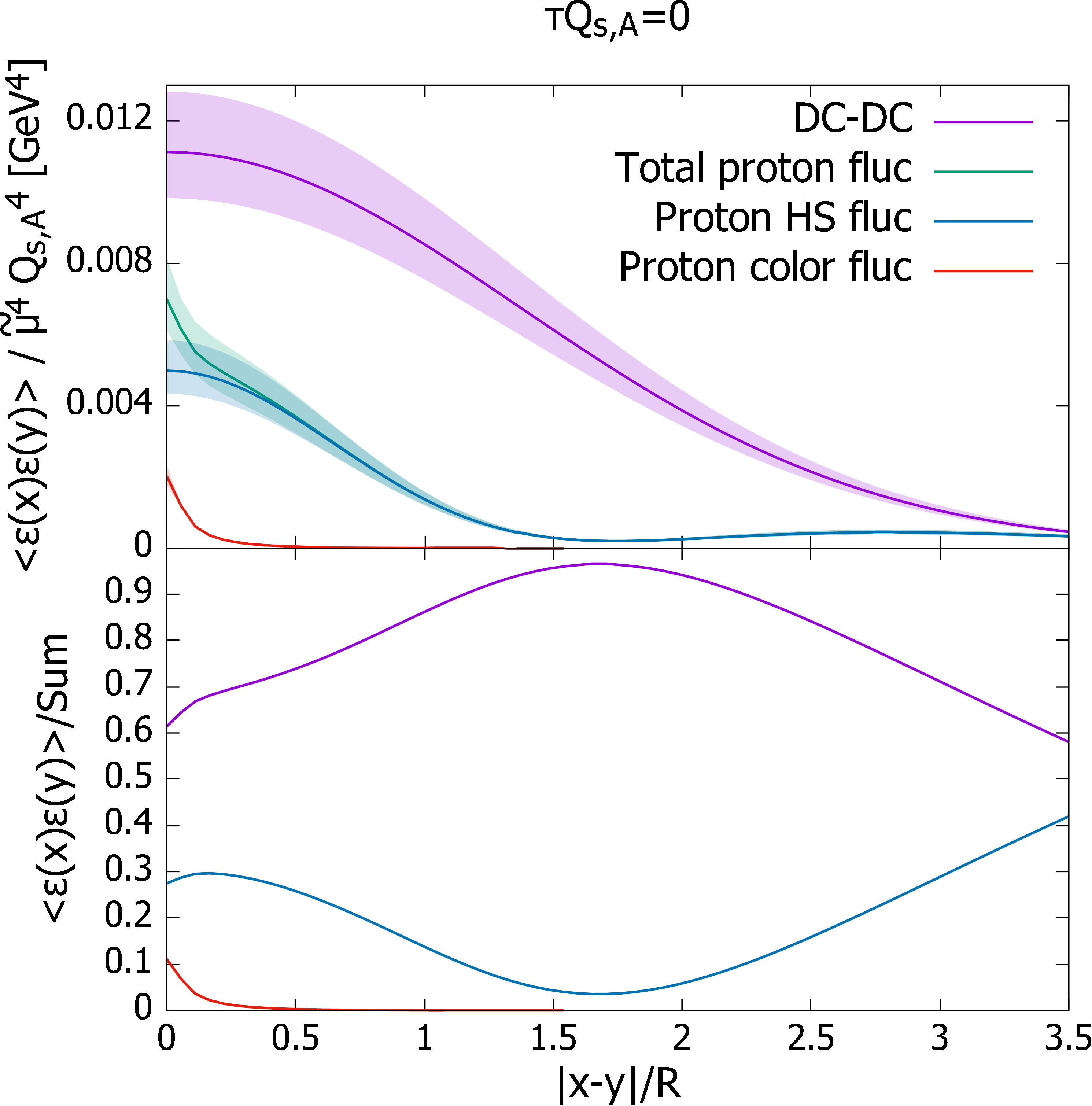}
\includegraphics[width=0.45\textwidth]{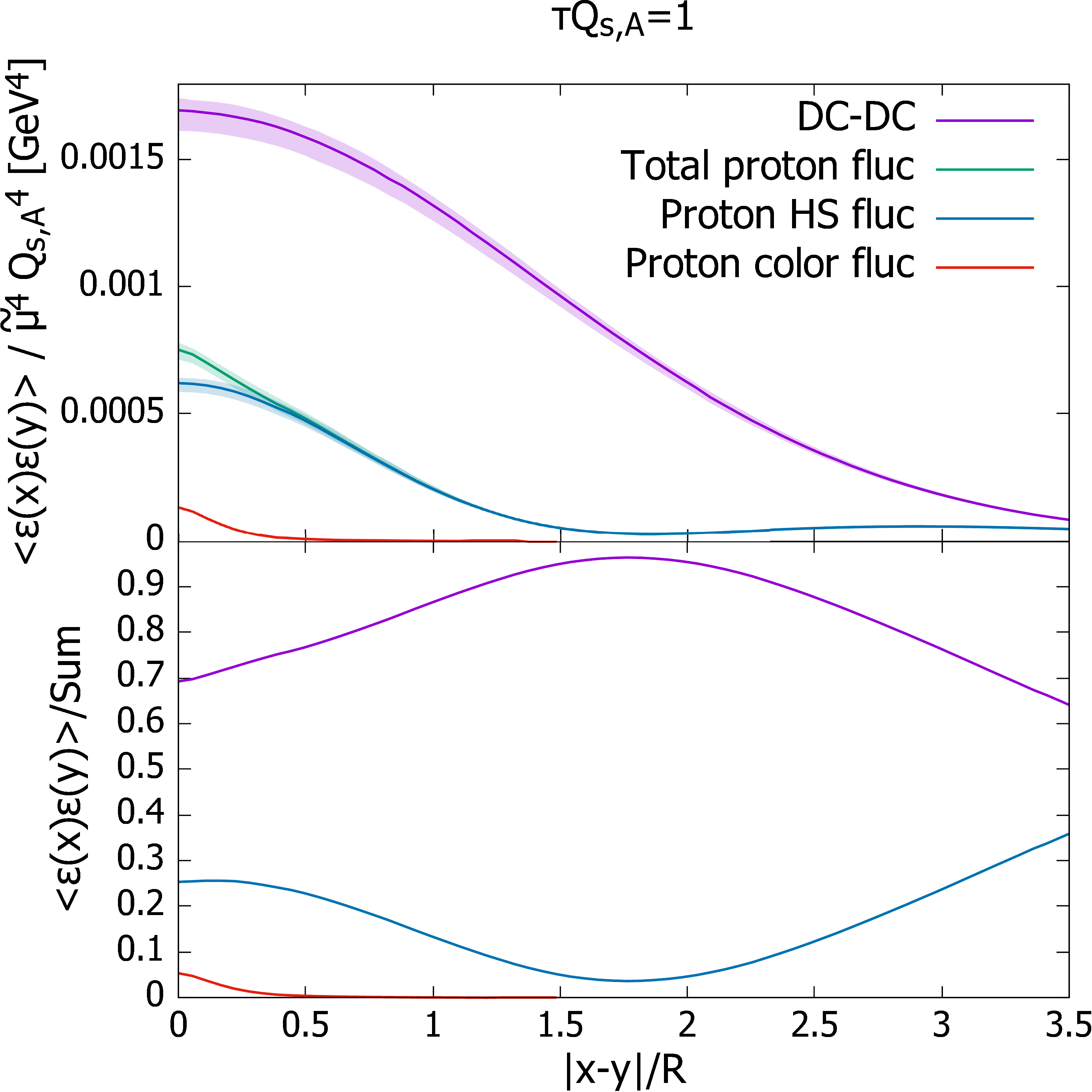}
\caption{Same as \figref{fig:2PointHSAndColor} for the non-local two-point function.}
\label{fig:2PointHSAndColor2}
\end{figure}

Let us now evaluate these expressions numerically. In Figs.\ \ref{fig:2PointHSAndColor}, \ref{fig:2PointHSAndColor2} we present plots of the two-point function of energy density $\langle\varepsilon(\tau,\xt)\varepsilon(\tau,\yt)\rangle$ at proper times $\tau = 0^+$ (left-hand side) and $\tau = 1/Q_{s}$ (right-hand side). In the top panels we show the fully disconnected contribution, the total proton fluctuations contribution, and the proton hot spot and color fluctuations separately. The bottom panels show the relative contribution of the separate parts with respect to the total. In \figref{fig:2PointHSAndColor} we set $\xt=\yt$ and show the two-point function as a function of the distance from the center of the proton $\Bt$, whereas in \figref{fig:2PointHSAndColor2} we set $\xt = -\yt$, start from $\xt=\yt=\Bt$ and let the coordinates move away from the center in such a way that $\left| \xt - \Bt \right|=\left| \yt - \Bt \right|$.

In Figs.\ \ref{fig:2PointHSAndColor}, \ref{fig:2PointHSAndColor2} we can see that the color fluctuations of the proton are short-range; their relative contribution approaches zero very fast as the separation distance grows. Also, the relative contribution of the color fluctuations seem to get slightly smaller as we evolve the system in $\tau$. As the eccentricity is mostly sensitive to the large distance behavior of the two-point function, this implies that the color fluctuations should not have a large effect on them (and even less so at larger values of $\tau$). The other part of the proton fluctuations is the hot spot fluctuation contribution. This is the dominant part of the full proton fluctuations except at small distances, where the color fluctuations give a sizable contribution. The proton fluctuations seem to go down as the separation increases, and then up again slightly. This is because the proton hot spot fluctuations consist of two different terms: a short-range one hot spot contribution (containing the function $F_2$) and a larger distance two hot spot contribution (containing the function $F_3$). The dip is due to the region where the dominant contribution changes. The relative contribution of the fully disconnected contribution dominates the two-point function up until the boundary of the proton. At long distances the proton fluctuations decrease at a slower rate and thus they become the dominant source of fluctuations.

\begin{figure}
\centering
\includegraphics[width=0.45\textwidth]{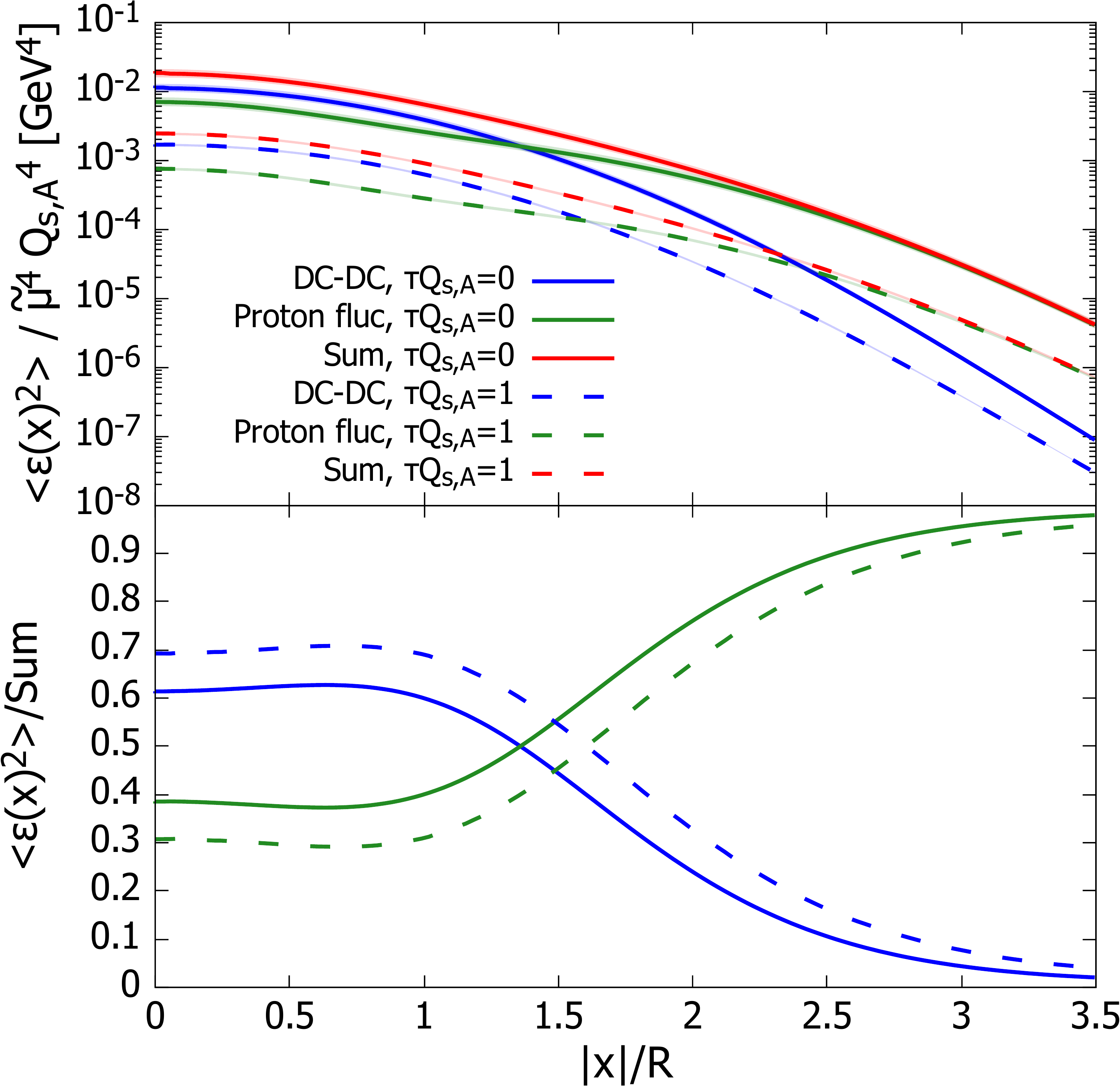}
\includegraphics[width=0.45\textwidth]{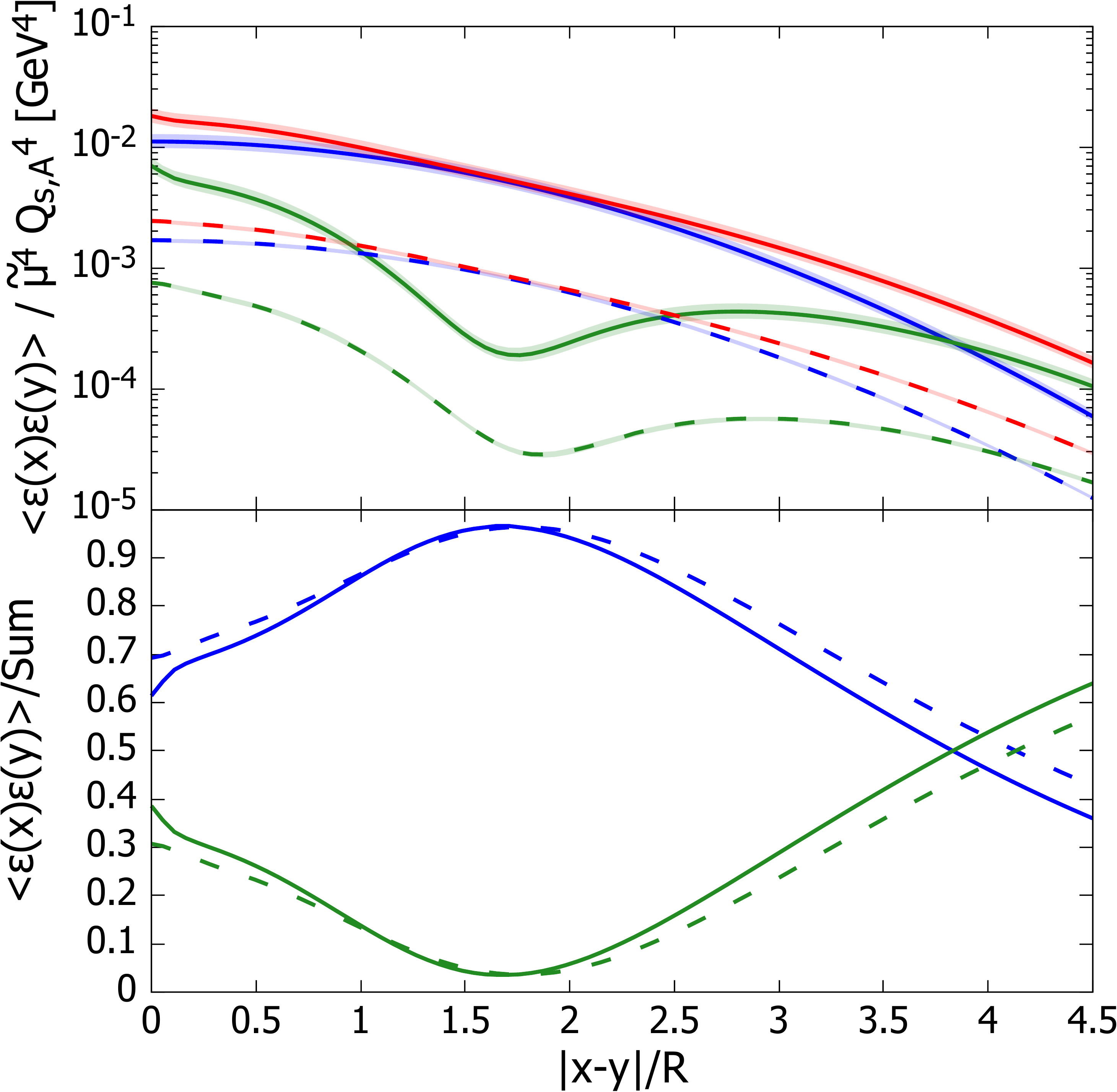}
\caption{The top panel shows the different contributions of the energy density two-point function at $\tau Q_s = 0$ and $\tau Q_s = 1$ plotted on a logarithmic scale. The bottom panels show the relative contribution of the different parts on a linear scale. The bands correspond to a variation of $C_0$ by $\pm20\%$. The legend in the plot on the left is also valid for the plot on the right.}
\label{fig:2PT0T1C0LogPlot}
\end{figure}

In the top panel of \figref{fig:2PT0T1C0LogPlot} we show the full two-point function, the proton fluctuations and the fully disconnected contribution on a logarithmic scale. The quantities are plotted at $\tau Q_{s}\!=\!0$ and $\tau Q_{s}\!=\!1$, illustrating the $\tau$-dependence of the different terms. Again, the bottom panel shows the relative contribution of the different parts at both proper times. Note that the right-hand side plots clearly show the aforementioned dip in the proton fluctuations. Also, one can see that these fluctuations become the dominant contribution of the two-point energy density, with the exact distance where this happens depending on $\tau$ (as we evolve the system further, the proton fluctuations start dominating at larger distances). This will have an effect on the eccentricities, which we will compute from these two-point functions in the following section.

\section{Eccentricities}\label{ecc}

Now we compute the eccentricities measuring the azimuthal correlations of the energy density field, as well as their response to $\tau$-evolution. Let us first summarize how we get the formulas for the two-particle eccentricities.

The eccentricity of an event quantifies how much the energy density field $(\varepsilon(\tau,\xt))$ deviates from an azimuthally symmetric configuration. It is defined as \cite{Blaizot:2014nia}:
\begin{equation} \label{eq:OnePEcc}
\varepsilon_n=\frac{\int d^2\xt \left| \xt - \Bt \right|^n \exp^{in\theta_{\xt-\Bt}} \varepsilon(\tau,\xt)}{\int d^2\xt \left| \xt - \Bt \right|^n \varepsilon(\tau,\xt)}.
\end{equation}
Note that the subindex $n$ allows to tell apart the symbols for eccentricities and energy density fields.
Note also that for the moment we are dealing with a singular event and thus have not done any averaging yet. The $\Bt$ appearing in \eqref{eq:OnePEcc} is the center of the event, defined as:
\begin{equation} \label{eq:CenterOfEvent}
\Bt = \frac{\int d^2 \xt\,\xt\,\varepsilon(\tau,\xt)}{\int d^2 \xt\,\varepsilon(\tau,\xt)}.
\end{equation}
We argue that $\Bt$ can be taken as the center of mass of the hot spot system, up to minor corrections stemming from the color charge fluctuations. This assumption lets us deal with the centering of the events analytically. In fact, as we are considering an infinitely large and homogenous nucleus, we can set $\Bt\!=\!0$ with no loss of generality.

Let us now write the energy density of a singular event as the sum of the average $(\left\langle\varepsilon(\tau,\xt)\right\rangle)$ and the fluctuation $(\delta \varepsilon(\tau,\xt))$:
\begin{equation} \label{eq:EDAvgFluc}
\varepsilon(\tau,\xt) = \left\langle \varepsilon(\tau,\xt) \right\rangle + \delta \varepsilon(\tau,\xt).
\end{equation}
Note that in this form it is evident that the average of the fluctuation of the energy density vanishes. For this reason, we argue that the simplest way to access the correlations caused by the fluctuations is to study the mean square of the eccentricity \eqref{eq:OnePEcc}, defined as:
\begin{equation} \label{eq:EccMeanSquare}
\varepsilon^{\prime}_n\{2\}^2 = \left\langle\frac{\int d^2\xt d^2\yt \left| \xt - \Bt \right|^n \left| \yt - \Bt \right|^n \exp^{in\left(\theta_{\xt-\Bt}-\theta_{\yt-\Bt}\right)} \varepsilon\left(\tau,\xt\right) \varepsilon\left(\tau,\yt\right)}{\int d^2\xt d^2\yt \left| \xt - \Bt \right|^n \left| \yt - \Bt \right|^n\varepsilon\left(\tau,\xt\right) \varepsilon\left(\tau,\yt\right)}
\right\rangle.
\end{equation}
Computing the average of this ratio analytically is not easy. Instead of doing that, we compute the approximation of this equation in the limit of small fluctuations. This approximation has been used previously in literature \cite{Blaizot:2014nia, Giacalone:2019kgg} and it reads:
\begin{equation} \label{eq:EccApprox}
\varepsilon^{\prime}_n\{2\}^2_{\text{Approx}} = \frac{\int d^2\xt d^2\yt \left| \xt - \Bt \right|^n \left| \yt - \Bt \right|^n \exp^{in\left(\theta_{\xt-\Bt}-\theta_{\yt-\Bt}\right)} \left\langle \varepsilon\left(\tau,\xt\right) \varepsilon\left(\tau,\yt\right) \right\rangle}{\int d^2\xt d^2\yt \left| \xt - \Bt \right|^n \left| \yt - \Bt \right|^n\left\langle \varepsilon\left(\tau,\xt\right)
\right\rangle \left\langle
\varepsilon\left(\tau,\yt\right) \right\rangle}.
\end{equation}
This approximation assumes small fluctuations and it is possible that our model does not fulfill this assumption with some parameter values. However, in Ref.\ \cite{Demirci:2021kya} the approximation \eqref{eq:EccApprox} was compared to the following expression:
\begin{equation} \label{eq:EccTwoPointDenominator}
\varepsilon_n\{2\}^2 = \frac{\int d^2\xt d^2\yt \left| \xt - \Bt \right|^n \left| \yt - \Bt \right|^n \exp^{in\left(\theta_{\xt-\Bt}-\theta_{\yt-\Bt}\right)} \left\langle \varepsilon\left(\tau,\xt\right) \varepsilon\left(\tau,\yt\right) \right\rangle}{\int d^2\xt d^2\yt \left| \xt - \Bt \right|^n \left| \yt - \Bt \right|^n\left\langle \varepsilon\left(\tau,\xt\right)
\varepsilon\left(\tau,\yt\right) \right\rangle},
\end{equation}
where the denominator has been replaced by a two-point function. Note that this expression looks more like a standard quantum mechanical expectation value, and the denominator is now sensitive to the fluctuations in the system. Through this comparison, it was found that \eqref{eq:EccMeanSquare} and \eqref{eq:EccTwoPointDenominator} differ very little from each other using the same parameters as we do, with the exception of the IR regulator (which in Ref.\ \cite{Demirci:2021kya} was taken as $m\!=\!0.22\text{ GeV}$). Also, this comparison was performed at $\tau\!=\!0^+$. As the value of $\tau$ increases, we found that the numerical analysis becomes very cumbersome, and thus made some choices to make it more reliable. Specifically, we used the approximation \eqref{eq:EccApprox} to compute the eccentricities and chose a larger value of the IR regulator $\left( m = 0.5\text{ GeV}\right)$.

\begin{figure}
\centering
\includegraphics[width=0.4\textwidth]{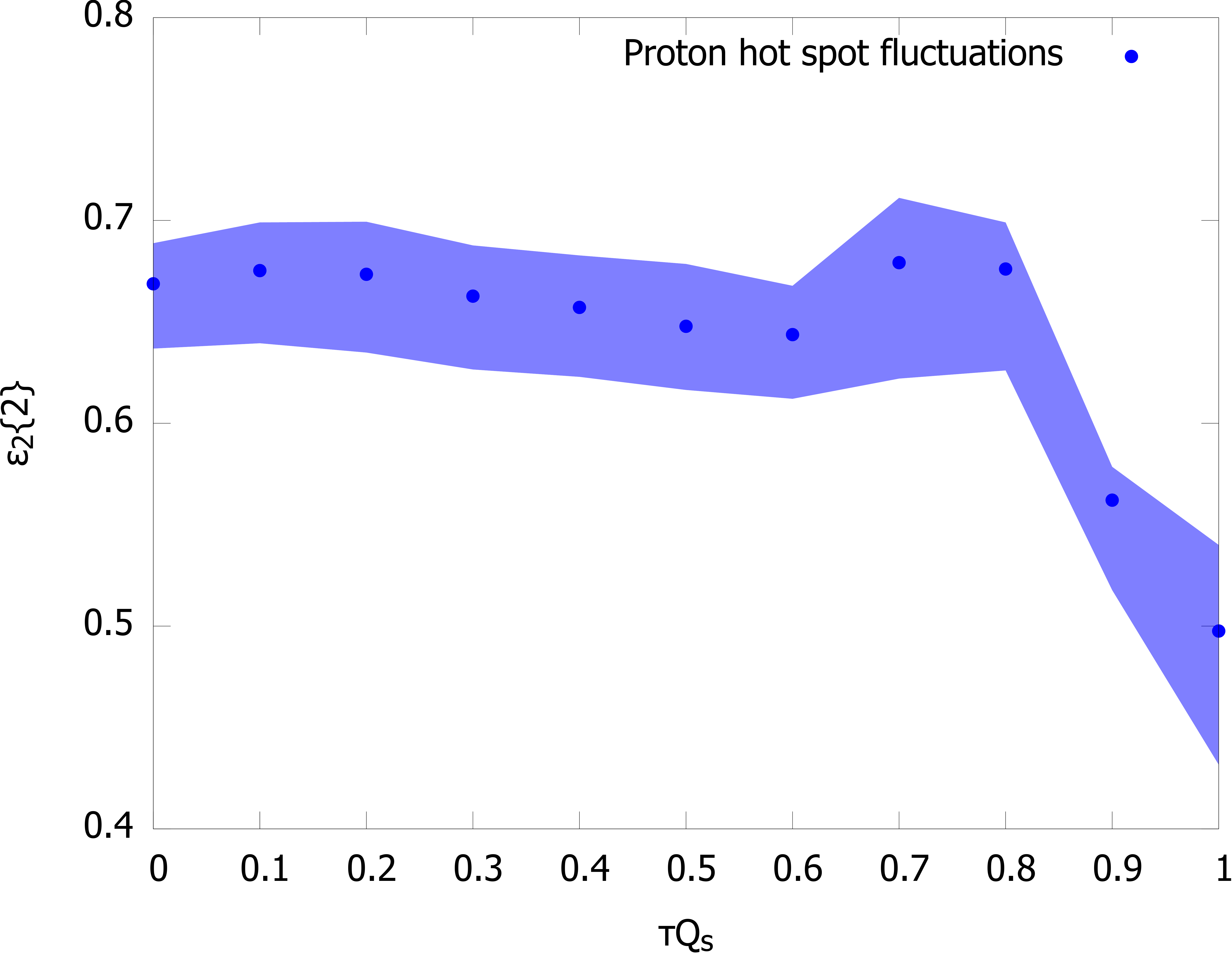}
\includegraphics[width=0.4\textwidth]{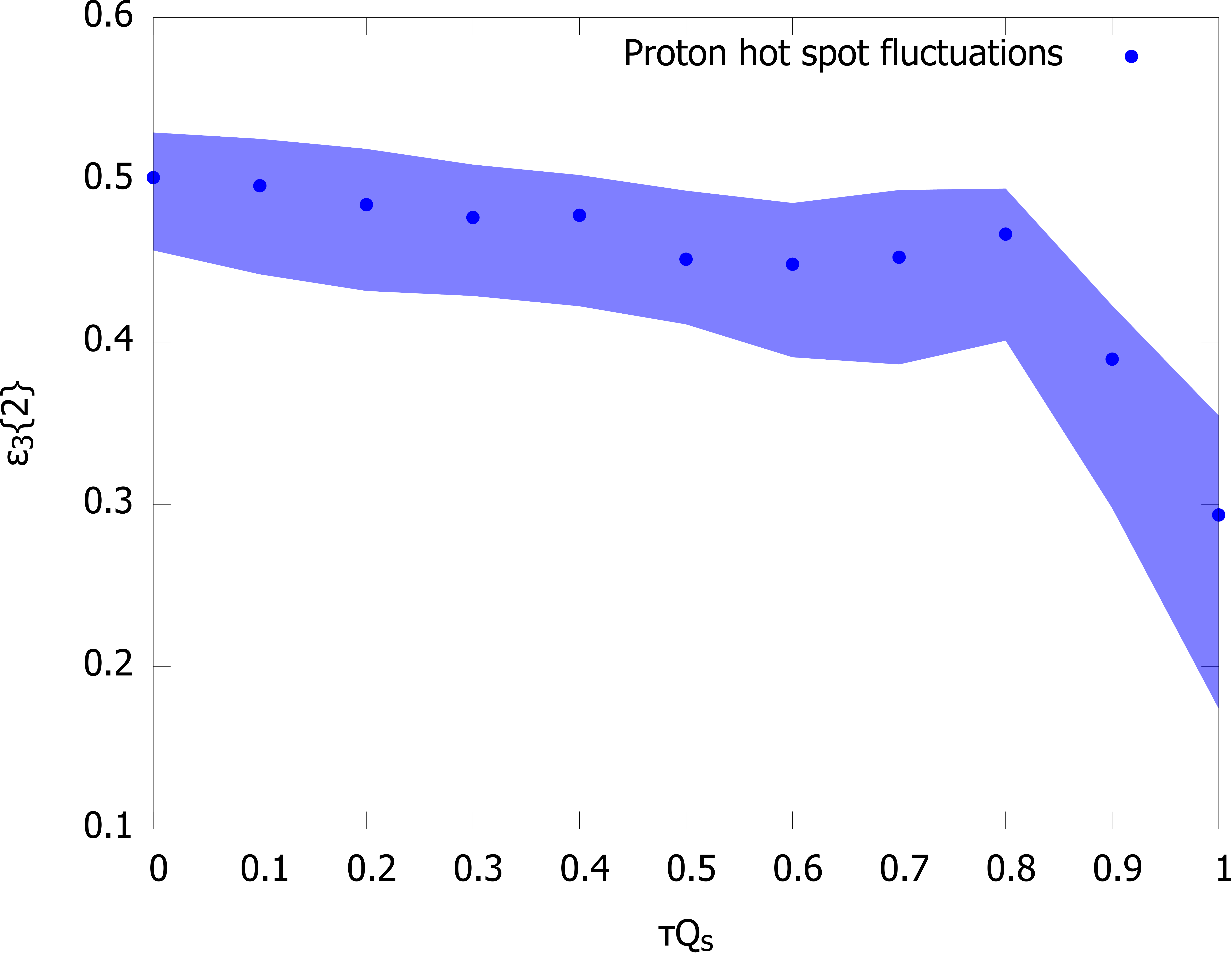} \\
\includegraphics[width=0.4\textwidth]{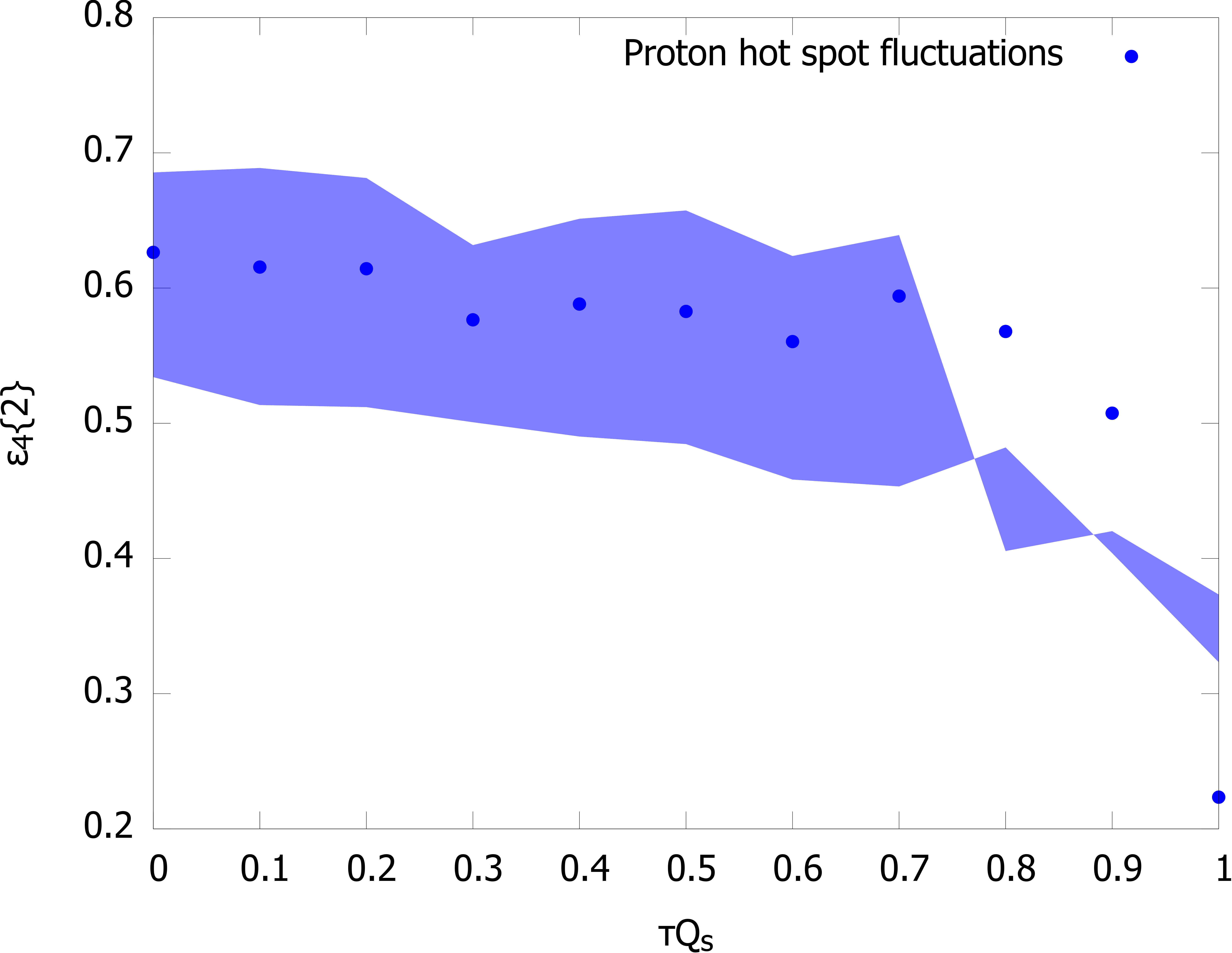}
\caption{The $\varepsilon_n \{2\}, n=2,3,4$ eccentricities as functions of $\tau Q_{s}$ with $Q_{s}\!=\!3 \text{ GeV}$. The confidence bands are found by varying the IR regulator $m$ by $\pm 20 \%$. The eccentricities are obtained taking into account only the hot spot fluctuations of the proton. This approximation is compared to the full result in Appendix \ref{app:AddRes}.}
\label{fig:EccNCFPlots}
\end{figure}

Our results for the $\tau$-dependent eccentricities with $n\!=\!2,3,4$ are shown in \figref{fig:EccNCFPlots}. These eccentricities only take into account the hot spot fluctuations of the proton. We expect that the color fluctuations only have a small effect in comparison (see Appendix \ref{app:AddRes} for more details). The confidence bands are found by varying $m$ by $ \pm 20\% $. Increasing the IR regulator makes the gluon field tails shorter and thus makes the hot spots more defined, increasing the eccentricity (whereas decreasing it does the opposite).

Looking at the confidence bands in \figref{fig:EccNCFPlots} we notice that the IR regulator dependence on the eccentricities remains nearly unchanged throughout the proper time evolution. This implies that whereas the UV regulator dependence was found to be small, the IR regulator plays an important role in the shape of the energy density field even when it is evolved in proper time. This further implies that one needs to properly fix the value of the IR regulator when using this kind of hot spot model to be able to constrain the values of the computed eccentricities (as well as any observables computed from them).

Another thing we notice is that whereas the initial $(\tau\!=\!0^+)$ eccentricities are generally quite high, there is a clear decreasing trend as a function of $\tau$. We do note however that the numerical instabilities begin to dominate as we go to higher $n$ and $\tau$, as can be seen from the $n\!=\!4$ plot. As the eccentricities are dominated by the long-range properties of the energy density field, we can try to understand the reason for the decreasing eccentricity by looking at the two-point function plots in Figs.\ \ref{fig:2PointHSAndColor}, \ref{fig:2PointHSAndColor2} and \figref{fig:2PT0T1C0LogPlot}. There we see that when $\tau$ becomes larger, the proton fluctuations start dominating the long distance behavior of the two-point function further away from the center. Here one should note that when computing the eccentricity, the two-point function is weighted by $\left| \xt - \Bt \right|^n \left| \yt - \Bt \right|^n$. From this it follows that the numerator of the eccentricity \eqref{eq:EccApprox} (given by the proton fluctuations) would decrease faster than the denominator (given by the fully disconnected contribution) as we evolve in $\tau$. This also means that the contribution of the center of the proton is suppressed whereas the edges are given more emphasis. This results in the observed decrease in eccentricity. However, having to work with four spatial coordinate parameters, it is difficult to plot the energy density in such a way that we could get the full picture of how the energy density behaves as a function of angles and relative distances.

\section{Conclusions}\label{fin}

In this work we study the effect of initial stage fluctuations on the early geometry of the system generated in high energy pA collisions. With this aim, we constructed a proper time-dependent model for the computation of energy density correlators in the dilute-dense limit. In this model we consider the proton at a moderately small-$x$, i.e.\ sufficiently small to permit the application of classical methods, but not so much that the proton becomes a homogeneous cloud of saturated gluons. This allows us to combine CGC initial conditions complemented with a hot spot model for the geometrical structure of the proton \cite{Demirci:2021kya} with a free field description of the subsequent evolution of the Glasma \cite{Lappi:2017skr}. It has been argued that the hot spot system is a natural way to include subnucleonic fluctuations, which are thought to be crucial in modeling the initial geometry of HICs \cite{Mantysaari:2020axf}. Moreover, free field propagation has been widely applied in the literature as an approximation of CYM evolution in the dilute-dense limit \cite{Dumitru:2001ux,Kovchegov:2005ss,Kovchegov:2005kn,McLerran:2016snu,Lappi:2017skr}, and, more recently, in the dense-dense case \cite{Guerrero-Rodriguez:2021ask}. This ensemble of elements allows for the analytical calculation of correlators at nonzero values of proper time.

We used our model to compute the energy density one- and two-point functions analytically. In this calculation we emphasized the contribution of the hot spot fluctuations, as those have been found to be the dominant source of eccentricity at $\tau\!=\!0^+$ in this particular hot spot model \cite{Demirci:2021kya}. More specifically, in the cited work it was established that at $\tau\!=\!0^+$ the proton-side fluctuations completely dominate over the nucleus color fluctuations with $N_q\!=\!3$. On these grounds, and because its computation becomes increasingly cumbersome as we evolve in $\tau$, we dropped this contribution completely. The fully connected contribution was also neglected, as it is expected to be doubly suppressed due to its sensitivity to the fluctuations of both proton and nucleus. Lastly, in Ref.\ \cite{Demirci:2021kya} it was found that the color fluctuation part of the full proton fluctuations yield a very small eccentricity when $N_q\!=\!1$. This result supports the argument that, as color fluctuations have a short range, one can expect them to have little effect on IR-dominated properties like the eccentricities. Overall, these considerations leave the hot spot fluctuations as the only relevant contribution in the computation of eccentricities. Said computation was the last part of this work, which we did considering values for the model parameters that we deemed reasonable (and that have in fact been used in previous studies). As a check of the validity of our approximation we also computed the eccentricities using the full proton fluctuations (see Appendix \ref{app:AddRes}). Although the observed effect is not large at lower values of $\tau$, at larger values ($\tau Q_s\!\gtrsim\!0.4$) the difference starts to be sizable. However, due to significant numerical instabilities kicking in around this point (and onwards), it is difficult to attribute said difference to a breakdown of our approximation. We thus find that the best way to reach reasonable conclusions is by removing the numerically problematic short-range contribution of the color charge fluctuations.

From our results we can see that, even though the eccentricities at early times are quite substantial, proper time evolution does decrease them. One can interpret this qualitatively as a consequence of the evolution of the different terms of the energy density correlators. When examining distances close to the center of the proton, the fully disconnected contribution to the energy density two-point function dominates over the proton fluctuation contribution. However, as we move further away from the center, the proton fluctuations start dominating instead. This effect is observable already at $\tau\!=\!0^+$, but, as we evolve the system in $\tau$, the region where the dominant contribution changes is pushed further away from the center. This is particularly relevant for the calculation of eccentricities, which emphasize the edges of the energy density field while suppressing the center. Even though we limited ourselves to study only two coordinate combinations of the two-point function, this effect quite intuitively explains the reduction in eccentricity with increasing proper time. We also found that the large IR regulator dependence of the initial eccentricity seems to persist through the proper time evolution.

The results of this work give analytical insight on the workings of free field evolution with a hot spot-based initial condition applied to the description of the earliest stages of pA collisions. We want to emphasize that the assumptions at the base of our model are not expected to be valid throughout the whole pre-thermalization phase, as the expansion of the medium will eventually lead to the breakdown of the classical field description. A more complete analysis would thus require an intermediate theory that matches CGC with hydrodynamics, with kinetic theory being a strong contender for this role \cite{Kurkela:2014tea,Kurkela:2015qoa,Keegan:2016cpi,Kurkela:2018vqr}. Another way of refining this calculation would be to abandon the dilute-dense limit in order to include proton charge density fluctuations to all orders (as we do with the nucleus). We also did not implement some physical ingredients that have been used in previous works using hot spots. These include: $Q_s$ fluctuations \cite{Schenke:2012wb,Schenke:2012hg,Mantysaari:2017cni,Mantysaari:2022ypp,Kumar:2022aly}, repulsion between hot spots \cite{Albacete:2016pmp,Mantysaari:2022ffw} and hot spots within hot spots \cite{Kumar:2021zbn}. Another interesting addition to our model could be the inclusion of $x$-dependence through $x$-dependent hot spot size and number as in Ref. \cite{Kumar:2022aly}.

\section*{Acknowledgements}

The authors thank Tuomas Lappi for useful discussions. S.D. is supported by the Vilho, Yrjö and Kalle Väisälä Foundation, and the Center of Excellence in Quark Matter of the Academy of Finland, and Project No. 321840. P.G.-R. acknowledges financial support by the European research Council project ERC-2018-ADG-835105 YoctoLHC.

\appendix

\section{Some additional results} \label{app:AddRes}

\begin{figure}
\centering
\includegraphics[width=0.4\textwidth]{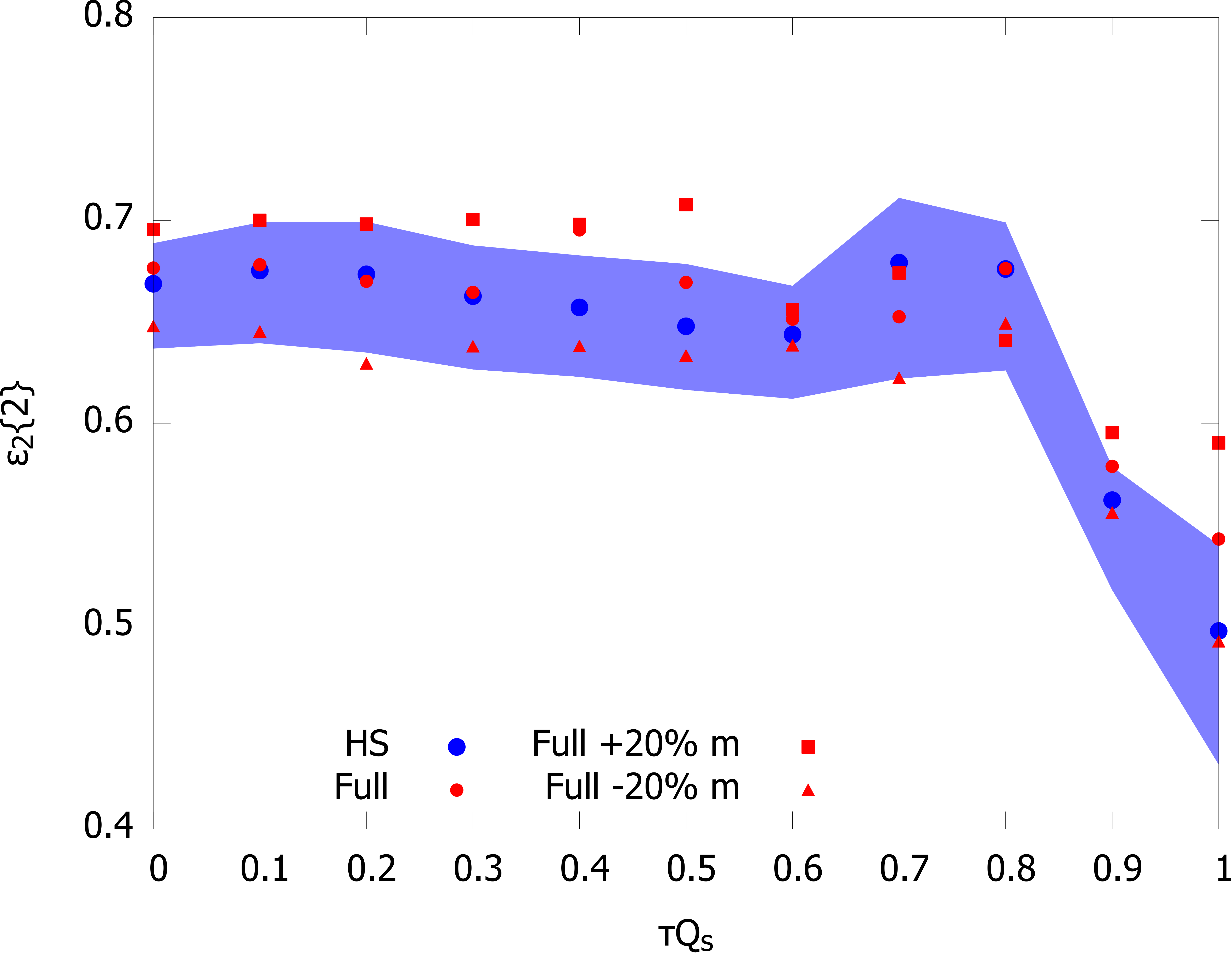}
\includegraphics[width=0.4\textwidth]{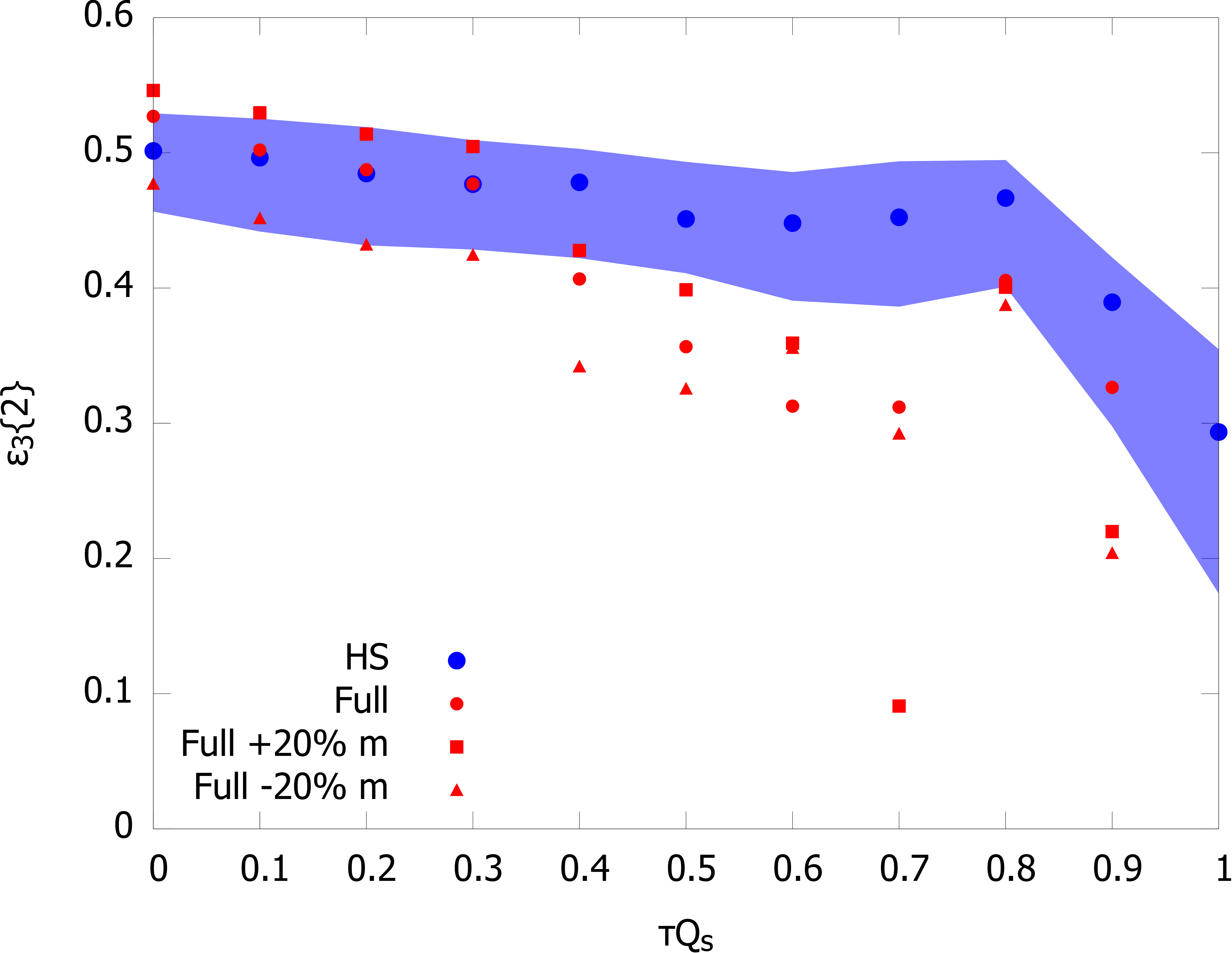} \\
\includegraphics[width=0.4\textwidth]{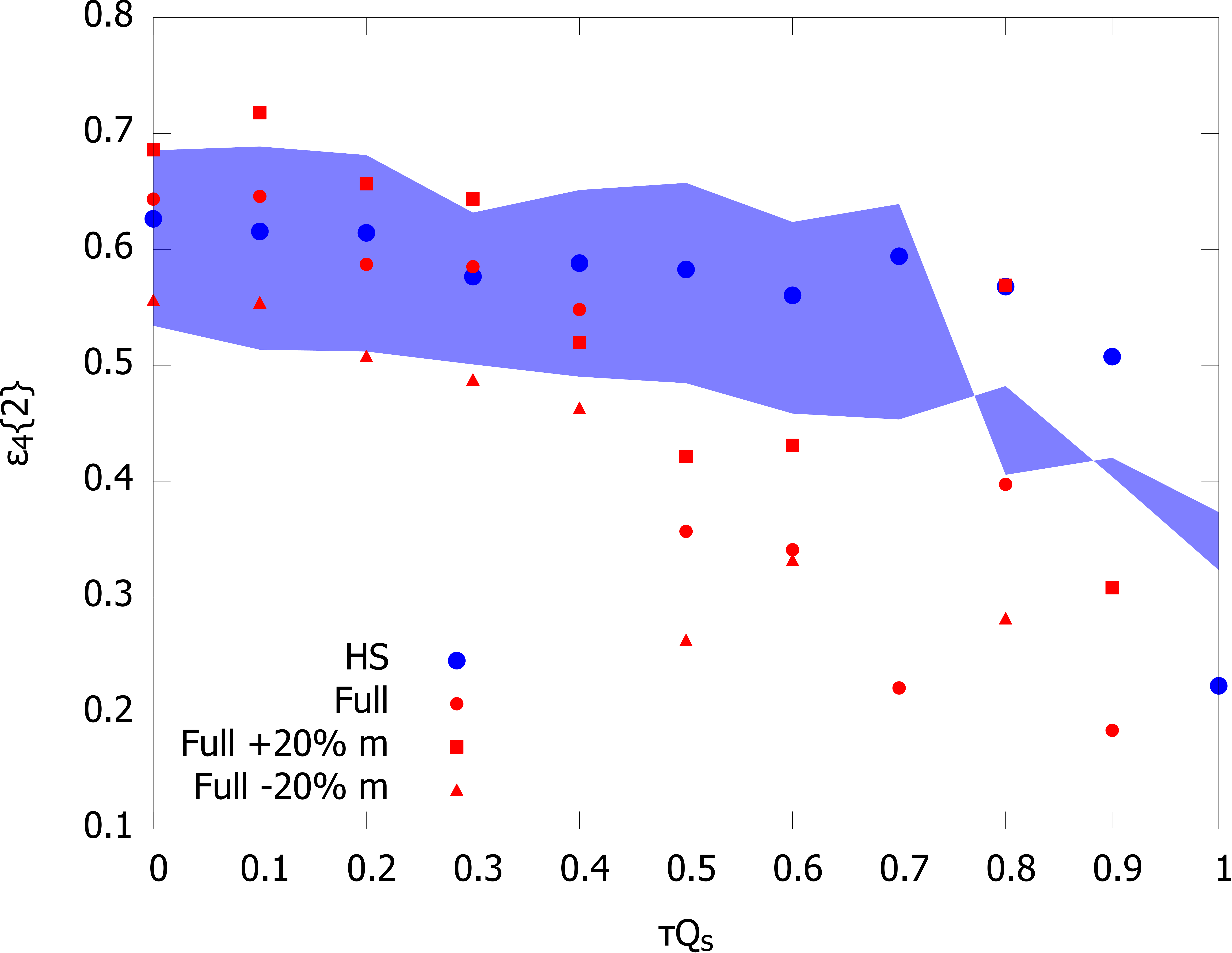}
\caption{The $\varepsilon_n \{2\}, n=2,3,4$ eccentricities as functions of $\tau Q_{s}$ with $Q_{s}=3 \text{ GeV}$. The confidence bands are found by varying the IR regulator $m$ by $\pm 20 \%$. Both the result from the approximation with only proton hot spot fluctuations and the result with all proton fluctuations are shown.}
\label{fig:EccFullAndNCF}
\end{figure}

In this appendix we present some additional plots that further explain why in the main text we focused on eccentricities computed by using only the hot spot fluctuation part of the proton fluctuations. \figref{fig:EccFullAndNCF} shows the results for both the full proton fluctuation and the approximation.

In \figref{fig:EccFullAndNCF} the full and the approximate result do coincide quite well at lower values of $\tau$. This makes us confident that our approximation works well and could give us a good estimate on what happens later in the evolution. As seen in \figref{fig:2PointHSAndColor}, the neglected proton color fluctuations are short-range, whereas the strong IR regulator dependence and the weak UV regulator dependence of the eccentricities suggest that they are most sensitive to the long distance behavior of the energy density. These observations reinforce our decision of dropping the color fluctuation part of the full proton fluctuations. The difference between the full result and the approximation observed at later stages of evolution is probably due to the color fluctuation contribution increasing the difficulty of the numerical integration, as at short range its relative contribution to the two-point function is still of the order of $\sim10\%$.

\bibliographystyle{JHEP-2modlong.bst}
\bibliography{main}

\end{document}